\documentclass[]{aastex631}
\usepackage{comment}

\usepackage{amsmath}
\usepackage{hyperref}
\usepackage{multirow}

\begin{document}
\title{ALMA Lensing Cluster Survey: Molecular Gas Properties of Line-Emitting Galaxies from a Blind Survey}

\author[0009-0000-5913-8555]{Kanako Narita}
\email{naritakn@g.ecc.u-tokyo.ac.jp}
\affiliation{Department of Astronomy, Graduate School of Science, The University of Tokyo, 7-3-1 Hongo, Bunkyo-ku, Tokyo 133-0033, Japan}
\affiliation{National Astronomical Observatory of Japan, National Institutes of Natural Sciences, 2-21-1 Osawa, Mitaka, Tokyo 181-8588, Japan}

\author[0000-0001-6469-8725]{Bunyo Hatsukade}
\affiliation{National Astronomical Observatory of Japan, 2-21-1 Osawa, Mitaka, Tokyo 181-8588, Japan}
\affiliation{Graduate Institute for Advanced Studies, SOKENDAI, Osawa, Mitaka, Tokyo 181-8588, Japan}
\affiliation{Department of Astronomy, Graduate School of Science, The University of Tokyo, 7-3-1 Hongo, Bunkyo-ku, Tokyo 133-0033, Japan}

\author[0000-0001-7201-5066]{Seiji Fujimoto}
\affiliation{David A. Dunlap Department of Astronomy and Astrophysics, University of Toronto, 50 St. George Street, Toronto, Ontario, M5S 3H4, Canada}
\affiliation{Dunlap Institute for Astronomy and Astrophysics, 50 St. George Street, Toronto, Ontario, M5S 3H4, Canada}

\author[0000-0003-3926-1411]{Jorge Gonz\'alez-L\'opez}
\affiliation{Instituto de Astrof\'isica, Facultad de F\'isica, Pontiﬁcia Universidad Cat\'olica de Chile, Santiago 7820436, Chile}
\affiliation{Millennium Nucleus for Galaxies (MINGAL)}
\affiliation{Las Campanas Observatory, Carnegie Institution of Washington, Ra\'ul Bitr\'an 1200, La Serena, Chile}


\author[0000-0002-4052-2394]{Kotaro Kohno}
\affiliation{Institute of Astronomy, Graduate School of Science, The University of Tokyo, 2-21-1 Osawa, Mitaka, Tokyo 181-0015, Japan}
\affiliation{Research Center for the Early Universe, Graduate School of Science, The University of Tokyo, 7-3-1 Hongo, Bunkyo, Tokyo 113-0033, Japan}

\author[0000-0001-6477-4011]{Francesco Valentino}
\affiliation{Cosmic Dawn Center (DAWN), Copenhagen, Denmark}
\affiliation{DTU Space, Technical University of Denmark, Elektrovej 327, DK2800 Kgs. Lyngby, Denmark}

\author[0000-0001-6653-779X]{Ryosuke Uematsu}
\affiliation{Department of Astronomy, Kyoto University, Sakyo-ku, Kyoto, Japan}

\author[0000-0003-3484-399X]{Masamune Oguri}
\affiliation{Center for Frontier Science, Chiba University, Chiba 263-8522, Japan}
\affiliation{Department of Physics, Graduate School of Science, Chiba University, Chiba 263-8522, Japan}

\author[0000-0002-5588-9156]{Vasily Kokorev}
\affiliation{Department of Astronomy, The University of Texas at Austin, Austin, TX, USA}

\author[0000-0002-8726-7685]{Daniel Espada}
\affiliation{Departamento de F\'{i}sica Te\'{o}rica y del Cosmos, Campus de Fuentenueva, Edificio Mecenas, Universidad de Granada, E-18071, Granada, Spain}
\affiliation{Instituto Carlos I de F\'{i}sica Te\'{o}rica y Computacional, Facultad de Ciencias, E-18071, Granada, Spain}

\author[0000-0003-1937-0573]{Hideki Umehata}
\affiliation{Institute for Advanced Research, Nagoya University, Furocho, Chikusa, Nagoya 464-8602, Japan}
\affiliation{Department of Physics, Graduate School of Science, Nagoya University, Furocho, Chikusa, Nagoya 464-8602, Japan}

\author[0000-0002-6610-2048]{Anton M. Koekemoer}
\affiliation{Space Telescope Science Institute, 3700 San Martin Drive,
Baltimore, MD 21218, USA}

\author[0000-0002-3405-5646]{Jean Baptiste Jolly}
\affiliation{Max Planck Institute for Exterrestrial Physics, Giessenbachstrasse 1 85748 Garching Germany}

\author[0000-0002-4622-6617]{Fengwu Sun}
\affiliation{Center for Astrophysics $|$ Harvard \& Smithsonian, 60 Garden St., Cambridge, MA 02138, USA}

\author[0000-0001-8183-1460]{Karina Caputi}
\affiliation{Kapteyn Astronomical Institute, University of Groningen, PO Box 800, 9700 AV Groningen, The Netherlands}
\affiliation{Cosmic Dawn Center (DAWN), Copenhagen, Denmark}

\author[0000-0003-0348-2917]{Miroslava Dessauges-Zavadsky}
\affiliation{Department of Astronomy, University of Geneva, Chemin Pegasi 51, 1290 Versoix, Switzerland}

\author[0000-0001-6920-662X]{Neil Nagar}
\affiliation{Universidad de Concepcion, Beltr\'{a}n Mathieu 253, Casila 160-C, Concepci\'{o}n, Chile}

\author[0000-0002-0498-5041]{Akiyoshi Tsujita}
\affiliation{Institute of Astronomy, Graduate School of Science, The University of Tokyo, 2-21-1 Osawa, Mitaka, Tokyo 181-0015, Japan}

\author[0000-0003-4193-9025]{Wei-Hao Wang}
\affiliation{Institute of Astronomy and Astrophysics, Academia Sinica, Taipei, 106216, Taiwan}

\begin{abstract}
We present results of a blind search for line-emitting galaxies using ALMA Lensing Cluster Survey data. 
We detected seven line emitters, one of which is [C\,{\sc ii}] at $z = 6.071$, four are CO at $z = 0.8$--1.1, and the remaining two are possibly CO or [C\,{\sc i}] within photometric redshift ranges.
Three of the four CO emitters are multiple images of the same galaxy. 
Compared to previous line-emitter searches in ALMA deep fields, our sample probes molecular gas masses $\sim$1 dex below the lower bound, thanks to gravitational lensing (typically $\mu \sim 4$, up to $\sim$30 in extreme cases). 
Most emitters are located in a region similar to normal star-forming galaxies in the star formation rate (SFR) versus molecular gas mass plane. 
To reduce dependence on SFR and stellar mass, we analyzed the molecular gas fraction and depletion timescale as a function of distance from the star-formation main sequence. 
We found that most emitters broadly follow the scaling relations from previous studies, consistent within the intrinsic scatter.
In addition, we serendipitously detected the CH $N = 1$, $J = 3/2 \rightarrow 1/2$ $\Lambda$-doublet transition from one CO emitter at $z = 1.142$, representing the first detection of CH from an individual galaxy at cosmological distances through a blind survey. 
The CH/CO column density ratio of $\sim$$10^{-4}$ is comparable to that of local AGN-host galaxies, suggesting that CH traces molecular gas associated with AGN activity, possibly irradiated by X-rays.
\end{abstract}


\section{Introduction}
\label{sec:intro}
It has been shown that the cosmic star-formation rate density 
rises from early times, peaks at z $\sim$ 1–3, and subsequently declines toward the present day (e.g., \citealt{annurev:/content/journals/10.1146/annurev-astro-081811-125615}; \citealt{Bouwens_2015}, and references therein).
This cosmic evolution could be driven by changes in the balance of gas supply, mainly through continuous accretion from the intergalactic medium and minor mergers \citep{2005MNRAS.363....2K, 2009Natur.457..451D}, as well as by variations in the star formation efficiency, or a combination of both.
The molecular gas content of distant galaxies has been mainly estimated via the CO line (e.g., \citealt{1987ApJ...319..730S}; \citealt{Tacconi_2013}) and the [C\,{\sc ii}] line (e.g., \citealt{2018MNRAS.481.1976Z, 2025A&A...693A.119C}), 
and also the dust continuum emission (e.g., \citealt{2012ApJ...758L...9M, 2016ApJ...820...83S}) in the millimeter and submillimeter regimes.
Despite substantial progress, the cosmic evolution of molecular gas mass density remains poorly constrained (e.g., \citealt{2013ARA&A..51..105C,2014ApJ...782...79W,2016ApJ...820...83S,2018ApJ...853..179T,2019ApJ...872....7R,2019ApJ...882..138D,2020ApJ...896L..21R,2020AJ....159..190L,2021A&A...646A..76L,2023ApJ...945..111B}). One major limitation is that many surveys have targeted galaxies preselected by their emission at other wavelengths (e.g., \citealt{Tacconi_2013}), 
which could introduce a bias in the, for example, star-formation rate (SFR), stellar mass, or dust obscuration of the targeted galaxies.
Furthermore, the limited survey volumes and sensitivities have hindered the detection of galaxies with faint line luminosities and small molecular gas masses, typically corresponding to detection limits of $\sim10^{10}\,M_\odot$ at $z\sim2$–3, such that galaxies with $M_{\mathrm{gas}} \lesssim \mathrm{few}\times 10^{9}\,M_\odot$ generally remain undetected, leading to a large uncertainty on the molecular gas mass density.

Recent blind surveys for (sub)millimeter line emitters with facilities such as the Plateau de Bure Interferometer (\citealt{2014ApJ...782...78D,2014ApJ...782...79W}), the Karl G. Jansky Very Large Array (VLA; \citealt{2018ApJ...864...49P,2019ApJ...872....7R,2020ApJ...896L..21R}), and Northern Extended Millimeter Array (NOEMA; \citealt{2023ApJ...945..111B}) have started to mitigate these biases.
Notably, deep blind surveys with the Atacama Large Millimeter/submillimeter Array (ALMA), such as the ALMA Spectroscopic Survey (ASPECS; \citealt{2019ApJ...882..138D,2020ApJ...902..110D}) in the Hubble Ultra Deep Field, have demonstrated that deeper observations lead to an increased number of detections, including sources with multi-wavelength counterparts that are missed in shallower surveys.
However, despite these deep surveys, sensitivity limitations prevent a full census of faint galaxies. The magnification induced by gravitational lensing allows us to detect fainter galaxies more efficiently (e.g., \citealt{2015A&A...577A..50D,2017ApJ...837...97L,2017A&A...608A.138G}). Previous studies have suggested that such faint galaxies — here broadly defined as those with lower molecular gas masses compared to previous CO-selected samples — may account for nearly 30\% of the cosmic molecular gas density (e.g., \citealt{2017A&A...605A..81D,Aravena_2019}), highlighting their significant but poorly understood role.

We exploit the strong lensing effect in the ALMA Lensing Cluster Survey (ALCS), an ALMA Cycle 6 Large Program (Project ID: 2018.1.00035.L; PI: K. Kohno) targeting 33 galaxy clusters with ALMA Band 6, to conduct a blind search for line emitters. By constructing a new sample that includes fainter, lower-molecular-gas-mass galaxies than those studied in previous blank-field surveys, we aim to systematically investigate their properties and to establish a foundation for future studies of cosmic molecular gas evolution. This enables us to examine whether the scaling relations established for brighter, more massive galaxies hold for these fainter populations, which remains an open question and is crucial for understanding the full picture of galaxy evolution across cosmic time.

The structure of this paper is as follows.
In Section \ref{sec:obs}, we briefly summarize the ALCS observations.  
In Section \ref{sec:res}, we present the method of line emitter searches and outline the spectral energy distribution (SED) fitting to estimate the photometric redshift (photo-$z$), SFR, and stellar mass.  
We derive the molecular gas mass in Section \ref{sec:mgas}.  
In Section \ref{sec:dis}, we compare the molecular gas properties of the detected line-emitting galaxies with those of other surveys.  
In Section \ref{sec:summary}, we describe the main conclusions of this work.  
Throughout this paper, we assume a standard $\Lambda$CDM cosmology with $H_0=70\ \mathrm{km\ s^{-1}\ Mpc^{-1}}$, $\Omega_{\Lambda}=0.7$, and $\Omega_{\mathrm{M}}=0.3$, , and adopt a \cite{2003PASP..115..763C} initial mass function (IMF). Where necessary, we converted SFRs given for a Salpeter IMF to the Chabrier IMF by multiplying by a factor of 0.63. Results based on a Kroupa IMF were left unchanged, since the difference with Chabrier is negligible. The stellar mass–SFR relation was also corrected accordingly to the Chabrier IMF \citep{annurev:/content/journals/10.1146/annurev-astro-081811-125615}.

\section{ALCS Observations and Data}\label{sec:obs}
The ALCS aims to map highly magnified regions in 33 massive galaxy clusters at 1.2 mm using Band 6 with a total survey area of 133 arcmin$^2$ \citep{2023pcsf.conf...16K, 2024ApJS..275...36F}. The cluster sample was selected from well-studied galaxy clusters featured in prominent HST treasury programs, including the Hubble Frontier Fields (HFFs; \citealt{2017ApJ...837...97L}), the Cluster Lensing and Supernova Survey with Hubble (CLASH; \citealt{Postman_2012}), and the Reionization Lensing Cluster Survey (RELICS; \citealt{2019ApJ...884...85C}). The ALMA observations were conducted between December 2018 and December 2019 in compact array configurations of C43-1 and C43-2, covering 26 clusters during Cycle 6 and seven clusters during Cycle 7. Two frequency setups were employed, centered at 259.4 GHz and 263.2 GHz with a bandwidth of 7.5 GHz, enabling a 15-GHz spectral scan across 250.0–257.5 GHz and 265.0–272.5 GHz. This approach expanded the survey volume for detecting line-emitting galaxies. For the clusters A2744, MACSJ0416, A370, and AS1063, observations were conducted in only one of the two frequency setups, as the observations for the other frequency setup were previously obtained during the ALMA Frontier Fields Survey 
\citep{2017A&A...608A.138G}. The details on the survey and data analysis are described in \cite{2024ApJS..275...36F}.

\begin{table}
\caption{Redshift coverage in the ALCS observations for emission lines expected for high-redshift galaxies.}
\label{table:archive-1a}
\centering
\begin{tabular}{ccc}
\hline\hline 
Line & Redshift Range & Rest Frequency \\
 & & (GHz) \\
\hline 
CO$(J=3-2)$ & 0.269--0.305, 0.343--0.383 & 345.796 \\
CO$(J=4-3)$ & 0.692--0.740, 0.790--0.844 & 461.041 \\
\sc[Ci]($^3P_1$ -- $^3P_0$) & 0.806--0.857, 0.790--0.844 & 492.16 \\
CO$(J=5-4)$ & 1.115--1.175, 1.238--1.305 & 576.268 \\
CO$(J=6-5)$ & 1.538--1.609, 1.685--1.766 & 691.473 \\
CO$(J=7-6)$ & 1.960--2.044, 2.133--2.227 & 806.652 \\
\sc[Ci]($^3P_2$ -- $^3P_1$) & 1.970--2.054, 2.143-2.237 & 809.34 \\
CO$(J=8-7)$ & 2.383--2.478, 2.580--2.687 &921.800 \\
\sc[N\,{\sc ii}]($^3P_1$ -- $^3P_0$) & 4.362--4.514, 4.674--4.845 & 1461.131 \\
\sc[C\,{\sc ii}]($^3P_{3/2}$ -- $^3P_{1/2}$) & 5.974--6.172, 6.381--6.602 & 1900.537 \\
\hline
\label{tab:line}
\end{tabular}
\vspace{0.5em}

\parbox{0.95\linewidth}{Note. The listed lines are selected based on their common detection in previous (sub)millimeter observations of high-redshift galaxies (e.g., \citealt{2013ARA&A..51..105C, 2019ApJ...882..138D, 2023ApJ...945..111B}). See Section~ \ref{subsec:id} for a detailed discussion. The CH line, although serendipitously detected in our sample (see Section~\ref{sec:ch}), is not included here as it was not among the originally targeted transitions in our blind search.}

\end{table}

\begin{figure}
\epsscale{0.5}
\plotone{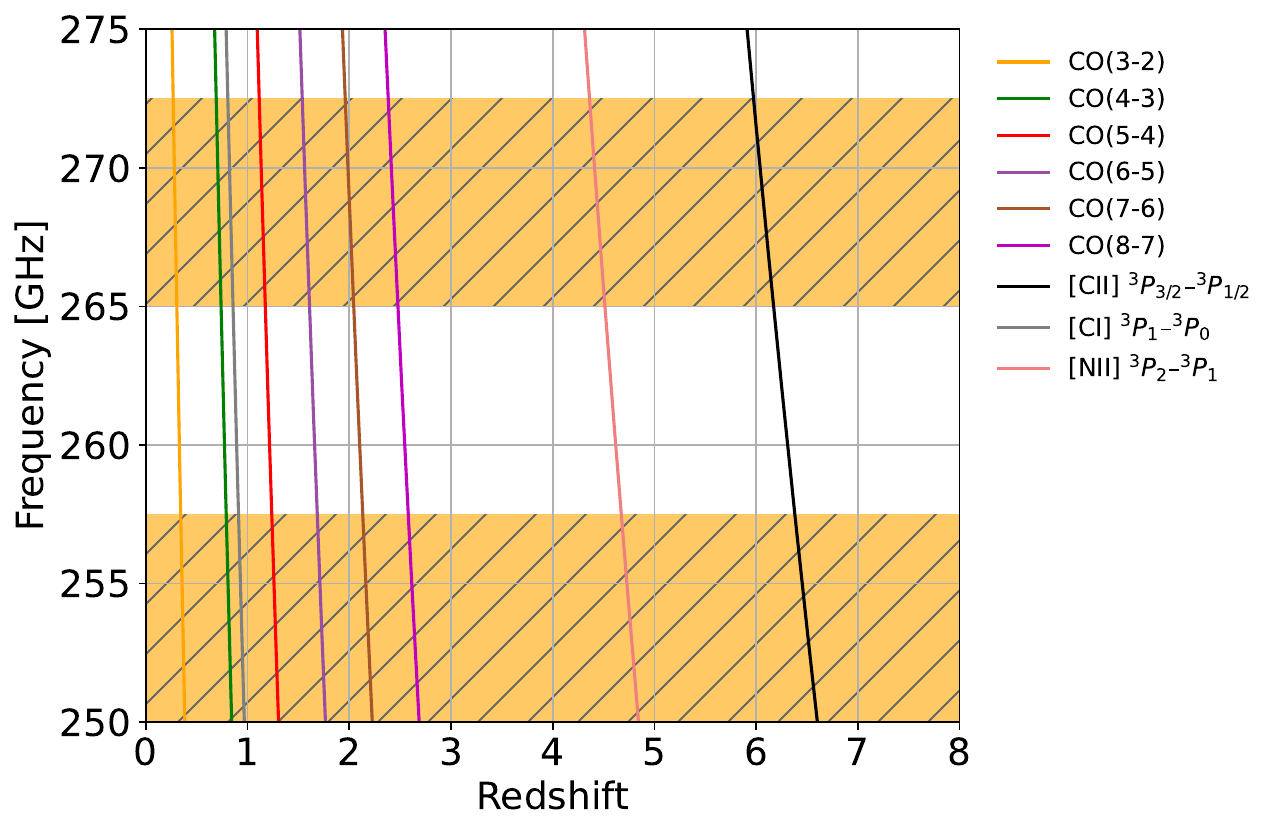}
\caption{The relationship between the observed frequency and redshift of emission lines expected for high-redshift galaxies. The shaded yellow region represents the frequency ranges observed by ALCS. 
}
\label{fig:redshift}
\end{figure}



\section{Emission Line Search} \label{sec:res}
\subsection{Method} \label{sec:emission:linesearch}
To search for line emitters, we used \texttt{LineSeeker} \citep{refId0}. This algorithm combines spectral channels using Gaussian kernels with varying widths. 
Kernel widths ranging from 60 to 1000 km s$^{-1}$ were used, where 60 km s$^{-1}$corresponds to the velocity resolution of the data cube, and 1000 km s$^{-1}$ reflects the typical upper limit of velocity widths observed in galaxies \citep{2013MNRAS.429.3047B}.
The channels are convolved with the corresponding Gaussian kernels, and high-significance features (i.e., voxels (three-dimensional pixels in the data cube) with S/N above the detection threshold) are identified channel by channel. The initial noise level is estimated and refined by excluding high-flux regions (i.e., pixels affected by line emission above the noise level), reducing the effect of bright emission lines. Voxels above a given signal-to-noise ratio (S/N) are stored for each convolution. The final list of line candidates is generated using Density-Based Spatial Clustering of Applications with Noise (DBSCAN) from the Scikit-learn library, and the S/N for each line candidate is selected as the maximum from all convolutions. We introduce fidelity ($F$), which expresses the confidence level of the emission line candidate, defined as,
\begin{equation}
\label{eq:d1}
F = 1-N_\mathrm{Neg}/N_\mathrm{Pos},
\end{equation} 
where $N_\mathrm{Neg}$ and $N_\mathrm{Pos}$ represent the number of negative and positive emission line candidates, respectively, with a given S/N value \citep{2016ApJ...833...67W}. The search was performed for data cubes with a velocity resolution of 60 km s$^{-1}$ and 150 km s$^{-1}$, resulting in 68 candidates with S/N $>$ 4 and $F > 0.9$ following the previous study \citep{refId0}. In practice, we restricted the range of spectral kernels used in the line search to physically plausible values, which effectively excludes candidates with unreasonably narrow or broad FWHM. 
Since our observations were conducted in a mosaic mode, the sensitivity is nearly uniform across the fields and the impact of the primary beam attenuation is negligible. 
In addition, all candidates were visually inspected in the spectra to remove spurious features.
Since the majority of the spectra of the line candidates are contaminated by continuum emission, we recreated data cubes by subtracting the continuum emission in the $uv$ plane with the \verb|uvcontsub| task of the Common Astronomy Software Applications (CASA) \citep{2007ASPC..376..127M, cite-key} by masking the channels where the preselected line emission candidates were detected. 
Then, dirty maps were created with a natural weighting. The typical rms noise levels were 0.85--1.7 mJy beam$^{-1}$ for the cubes with a velocity resolution of 60 km s$^{-1}$, and the typical spatial resolution was about 1 arcsec.

We measured the rms in the emission-free region of the maps for each channel in the continuum-subtracted cubes. 
We chose candidates that satisfy S/N $>$ 5 in the spectra, where the S/N refers to the integrated value obtained by summing the fluxes over the frequency channels covering the line, and S/N $>$ 4 in the continuum-subtracted velocity-integrated intensity (moment-0) maps, where the slightly looser threshold was adopted because these maps are more affected by correlated noise and residual imaging artifacts. 
In this way, we identified 16 emission line candidates, of which five are found to be cluster member galaxies.
Because the primary aim of this study is to investigate the molecular gas properties of line-emitting galaxies identified through a blind search (i.e., without any prior selection or pre-identified targets), we exclude the cluster member galaxies from the following analysis. 
The properties of the detected cluster member galaxies will be discussed in a separate paper. 

\subsection{Line identification} \label{subsec:id}
We identified detected line candidates focusing on molecular and atomic transitions frequently detected in high-redshift (sub-)millimeter spectroscopic surveys, including CO, [C\,{\sc i}], [C\,{\sc ii}], and [N\,{\sc ii}] (e.g., \citealt{2013ARA&A..51..105C, 2019ApJ...882..138D,2023ApJ...945..111B}). We note that higher-density tracers such as HCN or HCO$^+$ could in principle also be present in our frequency coverage, 
but they are expected to be much fainter than CO or [C\,{\sc ii}] and thus unlikely to be detected in our data. Figure~\ref{fig:redshift} provides a graphical representation of the transitions listed in Table~\ref{table:archive-1a}.
These transitions trace various phases of the interstellar medium (ISM) and are accessible within the ALCS frequency coverage for sources. The lower limit reflects the redshift of the galaxy clusters, while the upper limit is set by the frequency coverage of 212--272 GHz, beyond which the main CO and [C\,{\sc i}] transitions are shifted outside the band. CH is not included, as it has not been widely reported in high-redshift blind searches and is not among the commonly targeted transitions in previous studies. Nonetheless, we tentatively detected CH emission in M0553-C190, which we discuss further in Section ~\ref{sec:ch}.

\begin{figure}

\epsscale{0.9}
\plotone{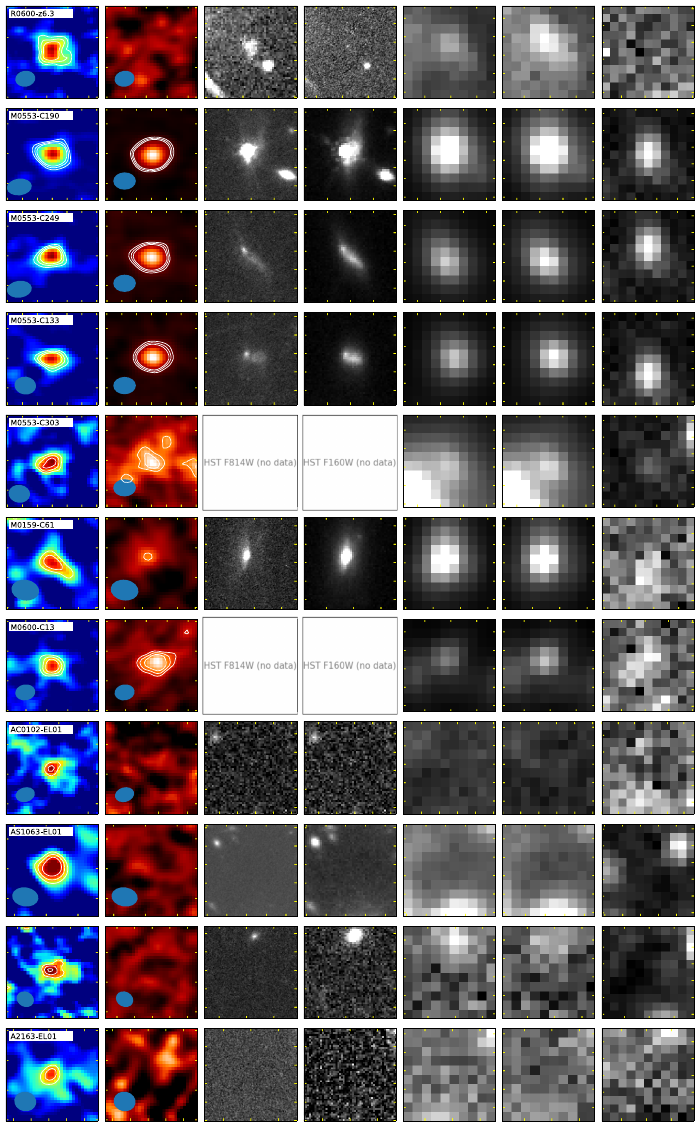}
\caption{Multi-wavelength image cutouts ($5'' \times 5''$) of the line emitters. 
From left to right: ALMA velocity-integrated intensity, ALMA 1.2 mm continuum, HST/F814W, HST/F160W, IRAC/ch1, and IRAC/ch2. 
The rightmost images are $100'' \times 100''$ cutouts of Herschel/SPIRE 350 $\mu$m.
The ALMA beam size is shown in the bottom left of the ALMA images.
The spectroscopic redshifts are shown in the leftmost images of the line emitters with counterparts. The contours indicate the intensity at $3\sigma$, $4\sigma$, and $5\sigma$ levels. R0600-C13 and M0553-C303 are not covered by HST imaging. M0553-C190, M0553-C249, and M0553-C133 correspond to multiple lensed images of the same galaxy.
}
\label{fig:img}
\end{figure}

To identify which molecular/atomic line corresponds to the detected line candidates, we investigate their optical to near-infrared (NIR) properties using the catalogs of HST, Spitzer/IRAC, and Herschel/SPIRE of the ALCS 33 fields \citep{Kokorev_2022}. 
Seven out of the 11 line emitter candidates have optical/NIR counterparts. 
In Figure \ref{fig:img}, we show the HST, IRAC, and Herschel images \citep{2022ApJ...932...77S}. We utilized the catalog of photometric redshifts \citep{Kokorev_2022} derived from optical–NIR SED analyses performed with the \texttt{easy} code \citep{Brammer_2008}.
Emission lines were identified based mainly on the \texttt{easy} SED results, with spectroscopic and photometric redshifts from the literature consulted when necessary. Details for individual sources are described in Appendix \ref{sec:indiv}.
 

We detected seven emission lines within the frequency ranges listed in Table~\ref{tab:line}.
Four of them are identified as CO transitions at $z = 0.8$–$1.1$, while one corresponds to [C\,{\sc ii}] emission at $z = 6.071$ (e.g., \citealt{Fujimoto_2021,2024ApJS..275...36F}).
For two sources, R0600–C13 and M0553–C303, both CO and [C\,{\sc i}] identifications are possible within their photometric redshift ranges.
Additionally, we identified four emission-line candidates without counterparts at other wavelengths.
To measure fluxes of the lines and continuum emission of the detected line emitters, 2-D Gaussian fitting using the CASA \texttt{imfit} task was conducted in the velocity-integrated intensity maps (moment 0) and the continuum maps. 
For resolved sources, we adopted the results from the \texttt{imfit} fitting. 
In contrast, for point sources and marginally resolved sources, we used the peak flux values measured in the moment-0 maps as an estimate of the integrated flux. 
Line widths were obtained by fitting a Gaussian function to the spectra extracted at the centroid positions of the lines. 

\begin{figure}
\epsscale{1.2}
\plotone{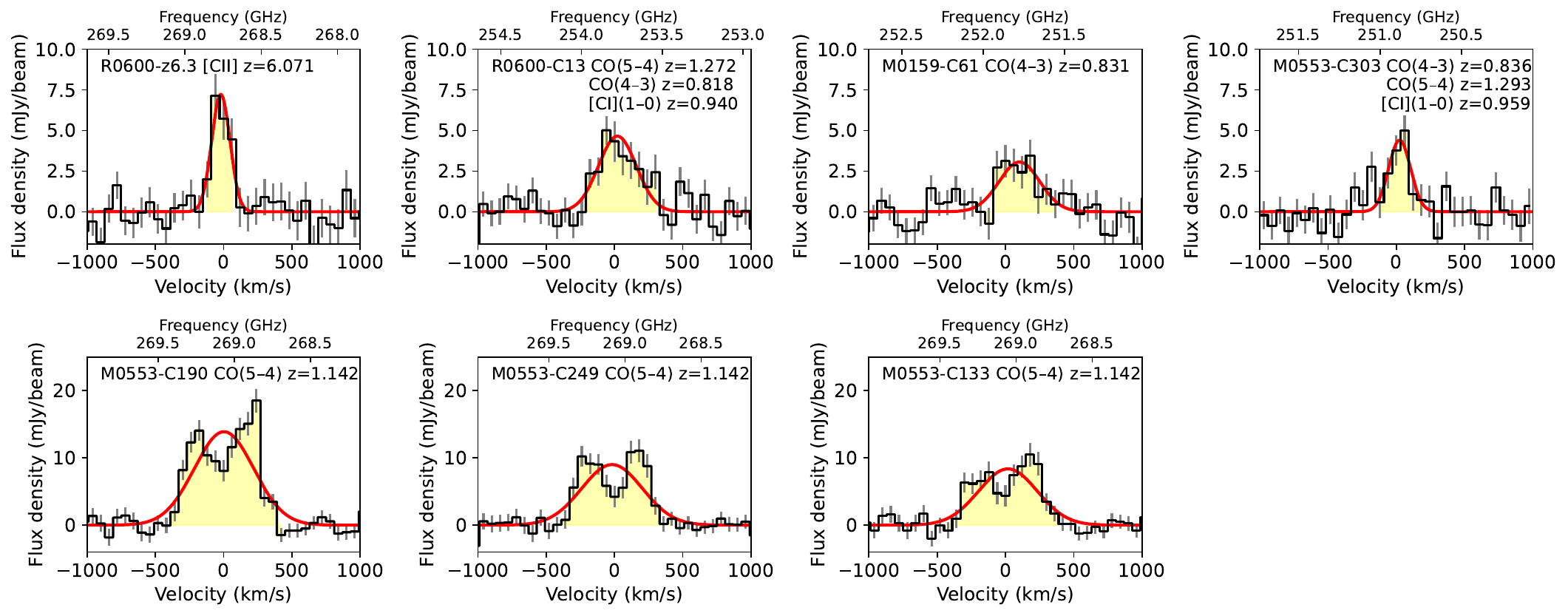}
\caption{Continuum-subtracted spectra of line emitters with counterparts. The error bars indicate the rms noise level measured in an emission-free region of the map for each channel. The spectra are shown with a velocity resolution of $60~\mathrm{km}~\mathrm{s}^{-1}$. The profiles are obtained by extracting the spectra at the location of the peak position identified in the moment 0 images. 
The yellow shaded area corresponds to the velocity range used to obtain the moment-0 images used in Figure \ref{fig:img}. The red curve shows the spectral fitting result.
}
\label{fig:spe1}
\end{figure}

\begin{figure}[htb!]
\epsscale{1.2}
\plotone{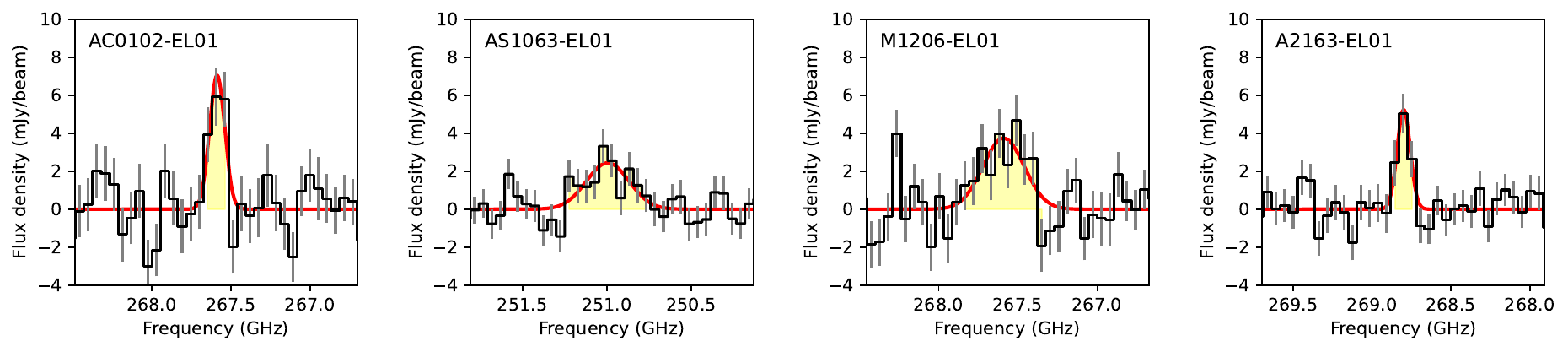}
\caption{Continuum-subtracted spectra of line emitter candidates without counterparts at other wavelengths. 
The error bars indicate the rms noise level measured in an emission-free region of the map for each channel. The spectra are shown with a velocity resolution of $60~\mathrm{km}~\mathrm{s}^{-1}$. The profiles are obtained by extracting the spectra at the location of the peak position identified in the moment-0 image. 
The orange shaded area corresponds to the velocity range used to obtain the moment-0 images used in Figure \ref{fig:img}. The red curve shows the spectral fitting result.
}
\label{fig:spe1-1}
\end{figure}

The observed properties of the line emitters are presented in Table \ref{table:emitters}. Figures \ref{fig:spe1} and \ref{fig:spe1-1} show the continuum-subtracted spectra of these line emitters with and without counterparts, respectively.
The four emission-line features that could not be robustly identified are treated as unidentified candidates throughout this work (See Section \ref{sec:ud}).

\begin{table}
\centering
\caption{Observed properties of the line emitters.}
\label{table:emitters}
\footnotesize
\begin{tabular}{ccccccccc}
\hline \hline
ID & Cluster & R.A. & Decl. & S/N & $F_{\rm line}$$^{*}$ & FWHM $^{\dagger}$ & $S_{\rm 1.2\,mm}$$^{\ddagger}$ & Note \\
   &       & (deg)& (deg) &  & (Jy km s$^{-1}$) & (km s$^{-1}$) & (mJy) & \\
\hline
R0600-z6.3 &RXCJ0600.1-2007 & 90.039843 & -20.136505 & 6.7  & 2.66$\pm$0.46 & 158$\pm$26 & -  & L21, F21, F24b, V24 \\
M0553-C190 &MACSJ0553.4-3342 & 88.366125 & -33.708529       & 20.5 &9.34$\pm$0.67$^{a}$ & 511$\pm$27 & 5.37$\pm$0.16 &  S21, F24a \\
M0553-C249 &MACSJ0553.4-3342& 88.365164 & -33.712218       & 17.7  & 5.87$\pm$0.66$^{a}$ & 522$\pm$42 & 3.57$\pm$0.14 &  S21, F24a \\
M0553-C133 &MACSJ0553.4-3342& 88.365857 & -33.704529 & 11.3  & 5.08$\pm$0.65$^{a}$ & 504$\pm$44 & 3.34$\pm$0.13 &  S21, F24a \\
M0553-C303 &MACSJ0553.4-3342& 88.383225 & -33.716837 & 5.9   & 0.85$\pm$0.17$^{a}$ & 188$\pm$38& 1.24$\pm$0.33 &  \\
M0159-C61  &MACSJ0159.8-0849& 29.960893 & -8.840993  &  6.2  & 1.19$\pm$0.28$^{a}$ & 351$\pm$73& 0.34$\pm$0.10 &      \\
R0600-C13  &RXCJ0600.1-2007 & 90.021143 & -20.164283 &  6.9  & 1.69$\pm$0.24 & 326$\pm$48 & 0.80$\pm$0.12 & F21      \\
\hline                                                                              
AC0102-EL01&ACTCLJ0102-49151& 15.799510 & -49.243519 & 6.5 & 2.21$\pm$0.58 & 126$\pm$29 &- & no counterpart \\
AS1063-EL01&AbellS1063& 342.186604& -44.539606 &  6.2  & 0.97$\pm$0.21$^{a}$ & 354$\pm$86 & - &no counterpart \\
M1206-EL01 &MACS1206.2-0847& 181.560791& -8.807667 &  6.5  & 3.19$\pm$0.70 & 343$\pm$82 & -&no counterpart \\
A2163-EL01 &Abell2163& 243.941370& -6.138139  &  6.6 & 2.68$\pm$0.63 & 110$\pm$16 &- &no counterpart \\
\hline
\end{tabular}
\tablecomments{
$^{*}$ Observed velocity-integrated fluxes, not corrected for gravitational lensing magnification. 
$^\dagger$ Full width at half maximum of the emission line. 
$^\ddagger$ Observed 1.2 mm continuum flux density, also not corrected for lensing. 
$^{a}$ Sources that are point-like or only marginally resolved. 
References in the “Note” column indicate previous studies in which the same object has been reported or discussed.\\
F21 \citep{Fujimoto_2021}, L21 \citep{2021MNRAS.505.4838L}, 
S21 \citep{2021ApJ...908..192S},
F24a \citep{2024ApJS..275...36F}, 
F24b \citep{2024arXiv240218543F},
V24 \citep{2024A&A...685A.138V}
}
\end{table}

\section{Physical properties of the line emitters}\label{sec:mgas}
\subsection{Magnification from gravitational lensing} \label{subsec:magnification}
To correct for the magnification by gravitational lensing, we adopt the lens model of \texttt{glafic} as a fiducial model following \cite{2024ApJS..275...36F}. 
We also used other available models to evaluate the systematic uncertainty of magnification factors ($\mu$). 
We adopt systematic uncertainties in $\mu$ following Section 3.6 of \cite{2024ApJS..275...36F}. 
Owing to the availability of JWST observations, the multiple images of R0600-z6.3 have already been spectroscopically confirmed, providing strong constraints on the lens model through their positions, flux ratios, and morphologies. Given these precise constraints from JWST, we adopted the magnification factor reported by \cite{2024arXiv240218543F} instead of applying the method described above for R0600-z6.3. We note that the magnification estimates from the earlier model \citep{2024ApJS..275...36F} and the updated JWST-based model \citep{2024arXiv240218543F} are consistent within $\sim$10$\%$ (See \citealt{2024arXiv240218543F}), so the impact on our results is minimal.

\subsection{Molecular gas mass from the CO line}
The CO luminosities, $L'_{\rm{CO}}$ in K km s$^{-1}$ pc$^2$, were deduced following \cite{Solomon_1997}:
\begin{equation}
L_{\mathrm{CO}}^{\prime}=3.25 \times 10^7 \nu_{\mathrm{obs}}^{-2}(1+z)^{-3} D_{\mathrm{L}}^2 F_{\mathrm{CO}} \text {, }
\end{equation}
where $\nu_{\mathrm{obs}}$ is the observed frequency of the CO line in GHz, $D_{\mathrm{L}}$ is the luminosity distance at redshift $z$ in Mpc, and $F_{\mathrm{CO}}$ is the velocity-integrated CO line flux in Jy km s$^{-1}$.
The CO luminosities calculated for the transition CO($J$ $\rightarrow$ $J$ -- 1) 
should be converted to the ground transition CO($J=1-0$) assuming a line luminosity ratio, $r_{J1}=L_{\mathrm{CO}J \rightarrow J-1}^{\prime}/L_{\mathrm{CO}1-0}^{\prime}$. 
We use $r_{41}=0.31 \pm 0.06$, and $r_{51}=0.23 \pm 0.04$, from the results of previous observations of massive main-sequence (MS) galaxies at $z \sim 1.5$ \citep{daddi2015}.

The mass of molecular hydrogen gas ($M_\mathrm{gas}$) can be derived using the $L'_{\mathrm{CO}}$-to-H$_2$ conversion factor, $\alpha_\mathrm{CO}$, in $M_{\odot}$ (K km s$^{-1}$ pc$^{2}$)$^{-1}$ as follows, 
\begin{equation}
M_\mathrm{gas}= \alpha_\mathrm{CO} L'_{\mathrm{CO}(1-0)} 
\end{equation}

We adopted $\alpha_\mathrm{CO}$ of MS galaxies (e.g., \citealt{Aravena_2019}), 3.6 $M_{\odot}$ (K km s$^{-1}$ pc$^{2}$)$^{-1}$, which includes the contribution from helium.
The derived molecular gas masses are listed in Table \ref{tab:mol_data_all}. 
It should be noted that the derived molecular gas masses do not account for the CO-dark molecular gas component. In addition, for CO-based estimates, an uncertainty arises from the assumption of a line ratio between the observed mid-$J$ transitions and the ground-state $J = 1-0$ line. 


\subsection{Molecular gas mass from the [C\,{\sc ii}] line}
The [C\,{\sc ii}] line luminosity in $L_\odot$ can be written as follows \citep{Solomon_1997}, 
\begin{equation}
L_\mathrm{[CII]}=1.04 \times 10^{-3} F_\mathrm{[CII]}  \nu_\mathrm{rest} (1+z)^{-1} D_L^2
\end{equation}
where $\nu_\mathrm{rest}$ is the rest frequency of the observed [C\,{\sc ii}] line in GHz 
and $F_\mathrm{[CII]}$ is the integrated [C\,{\sc ii}] line flux in Jy~km~s$^{-1}$.
It has been proposed that the [C\,{\sc ii}] line is an effective tracer of molecular hydrogen gas mass (e.g., \citealt{2018MNRAS.481.1976Z}). The mass of molecular hydrogen gas can be derived using the $L_\mathrm{[CII]}$-to-H$_2$ conversion factor ($\alpha_\mathrm{[CII]}$) in $M_\odot$~$ L_\odot^{-1}$ 
as follows, 
\begin{equation} 
M_\mathrm{gas}= \alpha_\mathrm{[CII]} L_\mathrm{[CII]}. 
\end{equation}

We adopted $\alpha_{[\mathrm{CII}]}$= 31 $M_\odot$~$ L_\odot^{-1}$ for MS galaxies with a median absolute deviation of 0.2 dex obtained by \cite{2018MNRAS.481.1976Z,2020A&A...643A...5D}. 
The derived molecular gas mass is presented in Table~\ref{tab:mol_data_all}. 

\subsection{Molecular gas mass from the [C\,\textsc{i}] line}

The molecular gas mass can also be estimated from the atomic carbon [C\,\textsc{i}](\,$^3P_1 \rightarrow \,^3P_0$) line following, e.g., \cite{2013MNRAS.435.1493A}. The H$_2$ mass is given by
\begin{equation}
M_{\mathrm{H}_2} = 1375.8 \, D_{\mathrm{L}}^2 \, (1+z)^{-1} 
\left(\frac{X_{\mathrm{CI}}}{10^{-5}}\right)^{-1}
\left(\frac{A_{10}}{10^{-7} \,\mathrm{s}^{-1}}\right)^{-1}
Q_{10}^{-1} 
\left(\frac{S \Delta v}{\mathrm{Jy\,km\,s^{-1}}}\right),
\end{equation}
where $D_{\mathrm{L}}$ is the luminosity distance in Mpc, $z$ is the redshift, $S\Delta v$ is the velocity-integrated [C\,\textsc{i}] line flux in Jy km\,s$^{-1}$, $X_{\mathrm{CI}}$ is the [C\,\textsc{i}]/H$_2$ abundance ratio, $A_{10}$ is the Einstein coefficient for spontaneous emission of the [C\,\textsc{i}](1--0) line, and $Q_{10}$ is the excitation factor. 

We adopt $X_{\mathrm{CI}} = 3.2 \times 10^{-5}$ ($\log_{10}[{\rm C\,I/H_2}] = -4.8 \pm 0.2$), $A_{10} = 7.93 \times 10^{-8}$ s$^{-1}$, and $Q_{10}=0.49 \pm 0.02$ at $z \sim 1.2$ \citep[e.g.,][]{2018ApJ...869...27V}. The total gas mass including helium is then obtained by multiplying by a factor of 1.36:
\begin{equation}
M_{\mathrm{gas}} = 1.36 \, M_{\mathrm{H}_2}.
\end{equation}

The derived gas masses from [C\,\textsc{i}] are summarized in Table~\ref{tab:mol_data_all}. 
Uncertainties are propagated from the line flux, magnification factor, excitation factor $Q_{10}$, and the assumed abundance ratio.

\begin{table}
\caption{Line identification and physical properties of the line emitters}
\label{tab:mol_data_all}
\begin{tabular}{ccccccccc}
\hline\hline
Object & Line ID & $z_{\rm spec}$ & $\mu$ &
$\log M_{\rm gas}$ & $\log M_*$ & $\log{\rm SFR}$ &
$\log\mu_{\rm gas}$ & $\log\tau_{\rm dep}$ \\
 & & & &($M_\odot$) & ($M_\odot$) & ($M_\odot\,{\rm yr}^{-1}$) &  & (Gyr) \\
\hline
R0600-z6.3 & [C\,{\sc ii}] & 6.071 &$32.5^{+0.7}_{-0.8}$&$9.0^{+0.20}_{-0.00}$ & $8.7^{+0.3}_{-0.1}$ & $0.4^{+0.7}_{-0.6}$ & $0.4^{+0.7}_{-0.3}$ & $-0.4^{+0.7}_{-0.6}$ \\
M0553-C190 & CO(5--4) & 1.142 & $6.5 \pm 3.3 $ &$10.8\pm0.1$ & $10.7\pm0.1$ & $1.8\pm0.2$ &0.1 $\pm$ 0.1 & 0.0 $\pm$ 0.2 \\
M0553-C249 & CO(5--4) & 1.142 &$5.9 \pm 3.0 $&$10.6\pm0.1$ & $10.6\pm0.1$ & $1.8\pm0.2$ & $0.0\pm 0.1$ & 0.2 $\pm$ 0.2 \\
M0553-C133 & CO(5--4) & 1.142 &$4.4 \pm 1.8$&10.7 $\pm$ 0.1 & $10.9\pm0.1$ & $1.9\pm0.2$ & $-0.2\pm0.1$ & $-0.2^{+0.1}_{-0.2}$ \\
M0159-C61  & CO(4--3) & 0.831 &$2.4 \pm 0.9$ &$10.1\pm0.1$ & $11.3\pm0.1$ & $0.7^{+0.4}_{-0.3}$ & $-1.1\pm0.2$ & $0.4 \pm 0.3$ \\
\hline
\multirow{3}{*}{M0553-C303} & CO(4--3) & 0.838  &$1.5 \pm 0.6$&$10.2\pm0.2$ & $11.1\pm0.5$& $2.2^{+0.3}_{-0.4}$ & $-1.0^{+0.4}_{-0.5}$ & $-1.0\pm0.3$ \\
                            & CO(5--4)$^\dagger$ & 1.300 & $0.9\pm 0.4$&$10.7\pm0.2$&$11.4\pm 0.5$&$2.4^{+0.5}_{-0.3}$&  $-0.7 \pm$ 0.1& $-0.7 \pm$ 0.3\\
                            & [C\,{\sc i}](1--0)$^\ddagger$ & 0.960 &$1.7\pm0.7$&$10.5\pm0.3$&$11.1\pm 0.5$ &$2.1^{+0.5}_{-0.3}$ &  $-0.6 \pm$ 0.2& $-0.6 \pm$ 0.4\\
\hline                            
\multirow{3}{*}{R0600-C13}  & CO(5--4)           & 1.271 &$1.4 \pm 0.6$& $10.8\pm0.2$ & $11.1\pm0.5$ & $1.5\pm0.4$ & $-0.3^{+0.3}_{-0.4}$ & $0.3\pm 0.4$ \\
                            & CO(4--3)$^\dagger$ & 0.820 & $1.3\pm 0.5$ & $10.5\pm0.2$ &$11.1\pm 0.5$&$1.6\pm 0.4$& $-0.6^{+0.3}_{-0.4} $& $-0.0$ $\pm$ 0.4 \\
                            & [C\,{\sc i}](1--0)$^\ddagger$ & 0.940 & $1.4 \pm 0.6$ & $10.7\pm0.3$ &$11.1\pm 0.5$&$1.5\pm 0.4$& $-0.4 \pm$ 0.2& 0.2 $\pm$ 0.4\\
\hline
\end{tabular}
\tablecomments{
Molecular gas mass, stellar mass, and SFR are corrected for lensing magnification.
For sources with multiple plausible line IDs (marked with $^\dagger$ or $^\ddagger$), only the adopted line (top row) includes full derived quantities; 
alternative IDs list the corresponding gas mass at the assumed redshift.
}
\end{table}

\section{Discussion}\label{sec:dis}
\subsection{Comparison with ALMA Frontier Field Survey}
\cite{2017A&A...608A.138G} performed a search for emission lines in the ALMA observations of five clusters of the Hubble Frontier Field Survey \citep{2017ApJ...837...97L}, Abell2744, MACSJ0416.1-2403, MACSJ149.51+2223, Abell370, and AbellS1063, where ALCS observations were also conducted.
They reported 26 line candidates identified by \texttt{LineSeeker}, which was used in this study, whereas we detected one in the same cluster fields. 
The discrepancy in the number of detections could be attributed to differences in the adopted fidelity thresholds.
Only two of their line candidates (A370-EL01 and MACSJ0416-EL01) fulfill the fidelity criterion $F > 0.9$ as adopted in this study. 
One of the two line candidates, A370-EL01, is identified as A370-C31 in \cite{2024ApJS..275...36F}, a member galaxy of a cluster.
In fact, when comparing with the Hubble Frontier Field survey results \citep{2017A&A...608A.138G}, relaxing the threshold to $F>0.6$ does not increase the number of background galaxy detections in the overlapping cluster fields; instead, it only adds low-reliability, optically-dark candidates.
For this reason, we adopted a stricter fidelity criterion of $F > 0.9$ to ensure a robust sample of line emitters. Among the 68 sources that satisfy this fidelity threshold, even when applying the additional stringent S/N criterion described in Section~\ref{sec:emission:linesearch}, we found that (except for the CH line candidate discussed later) the sources that failed to meet the S/N cut did not have multi-wavelength counterparts. Therefore, relaxing the fidelity threshold would not increase the final number of secure detections. This supports the necessity of maintaining a high fidelity threshold to minimize contamination from spurious candidates. This conservative approach ensures that our sample is optimized for follow-up studies and robust statistical analyses.


\subsection{Why are most of the galaxies detected in this blind survey at a redshift around 1?}

We discuss why the redshifts of the line emitters obtained in this study are concentrated around $z \sim 1$, except for the [C\,{\sc ii}] emitter at $z = 6.071$. First, the cosmic evolution of the molecular gas mass density shows a clear peak at $z \sim 1$--2 (e.g., \citealt{2019ApJ...872....7R, 2020ApJ...902..110D}). At $z \sim 0.5$, the molecular gas mass density is lower by a factor of a few compared to $z \sim 1$--2, making CO line detections more challenging. In addition, about two-thirds of the ALCS cluster fields are located at $z \gtrsim 0.38$, where the CO(3--2) line does not benefit from gravitational lensing magnification. Most of the CO(3--2) line emitters at these redshifts are expected to be cluster member galaxies and are therefore not the main focus of this study. Moreover, the frequency coverage of the ALCS observations allows for the detection of CO lines up to $J = 5$ at $z < 1.5$ (Table \ref{table:archive-1a}). At $z \gtrsim 1.5$, higher-$J$ CO transitions ($J > 5$) enter the observable bands, but these lines are generally weaker, and their excitation conditions are not yet fully understood.
Previous studies (e.g., \citealt{daddi2015}; \citealt{2024A&A...685A.138V}; \citealt{2021ApJ...909...56L}; \citealt{2020ApJ...902..109B}) have shown that the CO spectral line energy distribution (SLED) of main-sequence galaxies typically peaks around $J \sim 5$. This suggests that lower-$J$ transitions are more favorable for detection, which is consistent with the observational preference at $z < 1.5$.
Taken together, these factors — the cosmic evolution of molecular gas mass density, the limited gravitational lensing effect in the ALCS cluster fields, the frequency coverage of the observations, and the CO SLED excitation conditions — collectively explain why most of the CO line emitters detected in this survey are concentrated around $z \sim 1$.


\subsection{Properties of the line emitting galaxies}
\begin{figure}
\epsscale{0.5}
\plotone{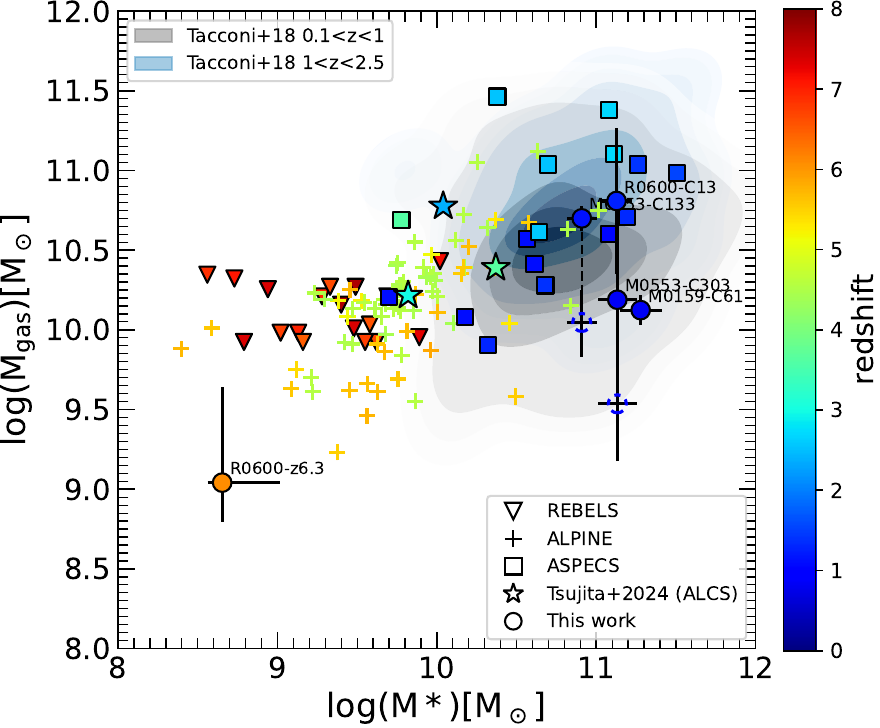}
\caption{Molecular gas mass vs stellar mass diagram for sources detected in ALCS (including three sources in \citealt{2024arXiv240609890T}), ASPECS \citep{Aravena_2019}, PHIBSS1/2 CO \citep{Tacconi_2013,2018ApJ...853..179T}, REBELS \citep{2024A&A...682A..24A}, and ALPINE \citep{2020A&A...643A...2B}. For sources with multiple possible line identifications, M0553–C303 and R0600–C13 are plotted assuming CO(4–3) and CO(5–4), respectively, following the identifications adopted in \cite{2024ApJS..275...36F}.}
\label{fig:Mmol_Mstar}
\end{figure}

The stellar mass ($M_*$) and SFR were derived from SED fitting by \cite{2024ApJ...965..108U} for the CO-detected galaxies and by \cite{2024arXiv240218543F} for the [C\,{\sc ii}]-detected galaxy (Table~\ref{tab:mol_data_all}). 
Note that the resolved SED modeling was detailed in \cite{2024A&A...686A..63G} and the IR part (especially for SFR$_{\rm{IR}}$, $M_{\rm{gas}}$ from dust, [C\,{\sc ii}], etc..) in \cite{2024A&A...685A.138V}.
The molecular gas depletion time ($\tau_{\rm dep} = M_{\rm gas}$/SFR) and molecular gas mass fraction ($\mu_{\rm gas} = M_{\rm gas}/M_*$) are derived in conjunction with the SED fitting results (Table~\ref{tab:mol_data_all}). 
M0553-C190, M0553-C249, and M0553-C133 are multiple images of the same source, and we use M0553-C133 as the representative in the following figures (Figure \ref{fig:Mstar_SFR}, Figure \ref{fig:Mmol_SFR} and Figure \ref{fig:Mmol_Mstar}), including the main-sequence diagram.
Figure \ref{fig:Mstar_SFR} shows the location of the ALCS sources in the SFR versus the stellar mass plane. 
The colors in the plot represent the respective redshifts. The solid lines show the empirical relationships between the stellar mass and SFR in MS galaxies at various redshifts deduced by \cite{2015A&A...575A..74S}. 
The ALCS line emitters at $z < 1.5$, except for M0159-C61, follow the main-sequence relation within the error range. 
The [C\,{\sc ii}] emitter at $z=6.071$ similarly follows the $z=6$ main sequence relation within the error range \citep{Fujimoto_2021}. 

\begin{figure}
\epsscale{0.5}
\plotone{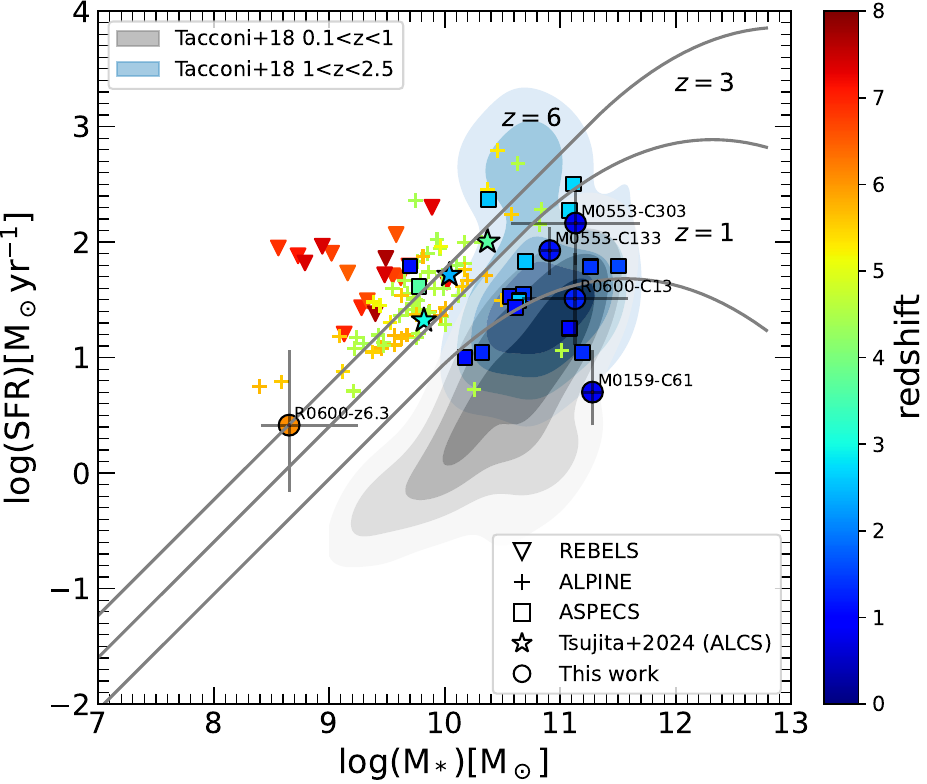}
\caption{Steller mass vs SFR estimated from the SED modeling diagram for ALCS sources (including three sources in \citealt{2024arXiv240609890T}), ASPECS sources \citep{Aravena_2019}, PHIBSS1/2 CO sources \citep{Tacconi_2013,2018ApJ...853..179T}, ALPINE sources \citep{2020A&A...643A...2B}, and REBELS sources \citep{2024A&A...682A..24A}.
Solid lines represent the main sequence of star-forming galaxies at $z = 1$, 3, and 6.}
\label{fig:Mstar_SFR}
\end{figure}

Figure \ref{fig:Mmol_SFR} shows the location of the ALCS line emitters in the SFR versus the molecular gas mass plane. 
The results of spectral line-scan observations of HST-dark sources found in ALCS \citep{2024arXiv240609890T} are also plotted with star marks. \cite{2024arXiv240609890T} reported that faint, triply lensed sources were detected in CO and [C\,{\sc i}] emission, confirming their redshifts of $z = 3.652$, 2.391, and 2.985. 
The majority of the ALCS sources reside in a similar region to normal star-forming galaxies with a gas depletion time of $\sim$1 Gyr, except for M0553-C133 and M0553-C303 with a shorter depletion time of $\sim$0.1 Gyr. 
For comparison, the results from the previous surveys are also plotted: 
ASPECS CO emitters at $ z \sim 1{-}3 $\citep{Aravena_2019} with squares, 
PHIBSS1/2 CO-detected sources \citep{Tacconi_2013,2018ApJ...853..179T} at $0.1 < z < 1$ with gray contours and at $1 < z < 2.5$ with light blue contours. 
We also plot the results of [C\,{\sc ii}] observations from the ALMA Large Program to INvestigate
[C\,{\sc ii}] at the Early Times (ALPINE; Dessauges-Zavadsky et al. 2020) and the Reionization Era Bright Emission Line
Survey (REBELS; Bouwens et al. 2022) with crosses and circles, respectively.
ALPINE is an ALMA Large Program in Cycle 5 that focuses on the observations of [C\,{\sc ii}] 158 $\mu$m line and FIR continuum emission for 118 UV-selected normal star-forming galaxies located in the main sequence. 
Most of the ALPINE galaxies have SFR $>$ 5 $M_{\odot}$~yr$^{-1}$ and $L_\mathrm{[CII]} > 5 \times 10^{7}$ $L_\odot$ \citep{2020A&A...643A...2B}. 
REBELS is an ALMA Large Program in Cycle 7 observing UV-selected star-forming galaxies at $z > 6.5$. 
About half of the REBELS sample consists of starburst galaxies with SFR $>$ 10 $M_\odot$ yr$^{-1}$ and $L_\mathrm{[CII]} > 2 \times 10^{8} L_{\odot}$. 
Compared to the results of the blind line search in ASPECS \citep{Aravena_2019}, our ALCS sample, thanks to gravitational lensing, probes molecular gas masses down to roughly 1 dex below the lower bound reached by ASPECS.
Although a direct comparison of median values is not straightforward because some sources have multiple plausible line identifications, the overall trend indicates that ALCS is sensitive to intrinsically fainter CO emitters at $z\sim1$ than those detected in ASPECS.
We note that the CO line ratios ($r_{J1}$) and conversion factors ($\alpha_{\rm CO}$) adopted here are calibrated for main-sequence (MS) galaxies.
If some of our sources instead follow starburst-like excitation conditions, they would exhibit higher CO excitation (e.g., $r_{51}\approx0.3$–0.4 compared to $r_{51}\approx0.2$ in MS galaxies) \citep{2020A&A...641A.155V} and a smaller $\alpha_{\rm CO}$ ($\sim0.8$ instead of $3.6~M_\odot\,{\rm (K \,km \,s^{-1} \,pc^{2})^{-1}}$) \citep{2013ARA&A..51..207B}.
Applying MS-based parameters to such systems would lead to a systematic overestimation of the molecular gas mass by up to an order of magnitude, depending on the transition used.
Therefore, the gas masses reported here should be regarded as upper limits, and future multi-line follow-up observations will be essential to refine these estimates.


\begin{figure}
\epsscale{0.5}
\plotone{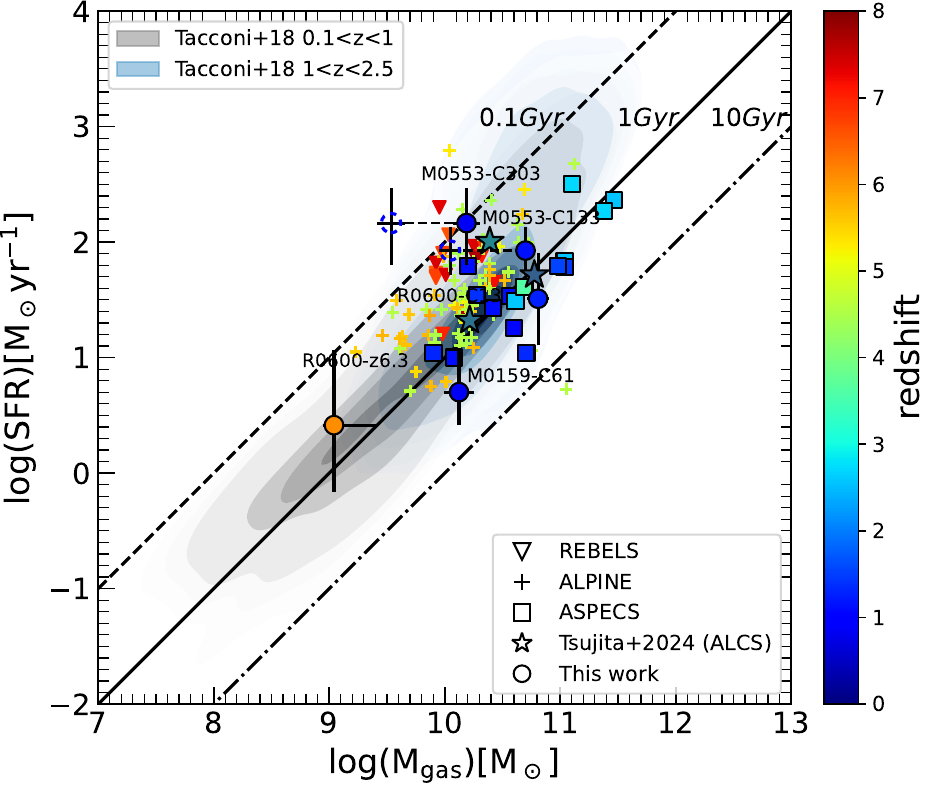}
\caption{Molecular gas mass vs SFR estimated from the SED modeling diagram for sources detected in ALCS (including three sources in \citealt{2024arXiv240609890T}), ASPECS \citep{Aravena_2019}, PHIBSS1/2 CO \citep{Tacconi_2013,2018ApJ...853..179T}, REBELS \citep{2024A&A...682A..24A}, and ALPINE \citep{2020A&A...643A...2B}. 
The dashed, solid, and dot-dash lines show the gas depletion time ($\tau_\mathrm{dep}$) of 0.1, 1, and 10 Gyr, respectively. For sources with multiple possible line identifications, M0553–C303 and R0600–C13 are plotted assuming CO(4–3) and CO(5–4), respectively, following the identifications adopted in \cite{2024ApJS..275...36F}.}
\label{fig:Mmol_SFR}
\end{figure}

Figure \ref{fig:Mmol_Mstar} shows the locations of the ALCS sources on the molecular gas mass versus stellar mass plane.
The ALCS sources at $z \lesssim 1$ appear broadly consistent with the PHIBSS galaxies at similar redshifts \citep{Tacconi_2013,2018ApJ...853..179T}, although their large uncertainties in gas mass prevent us from drawing firm conclusions about possible systematic offsets.


\subsection{Cosmic evolution of gas fraction and depletion time}
\begin{figure}
\epsscale{1}
\plotone{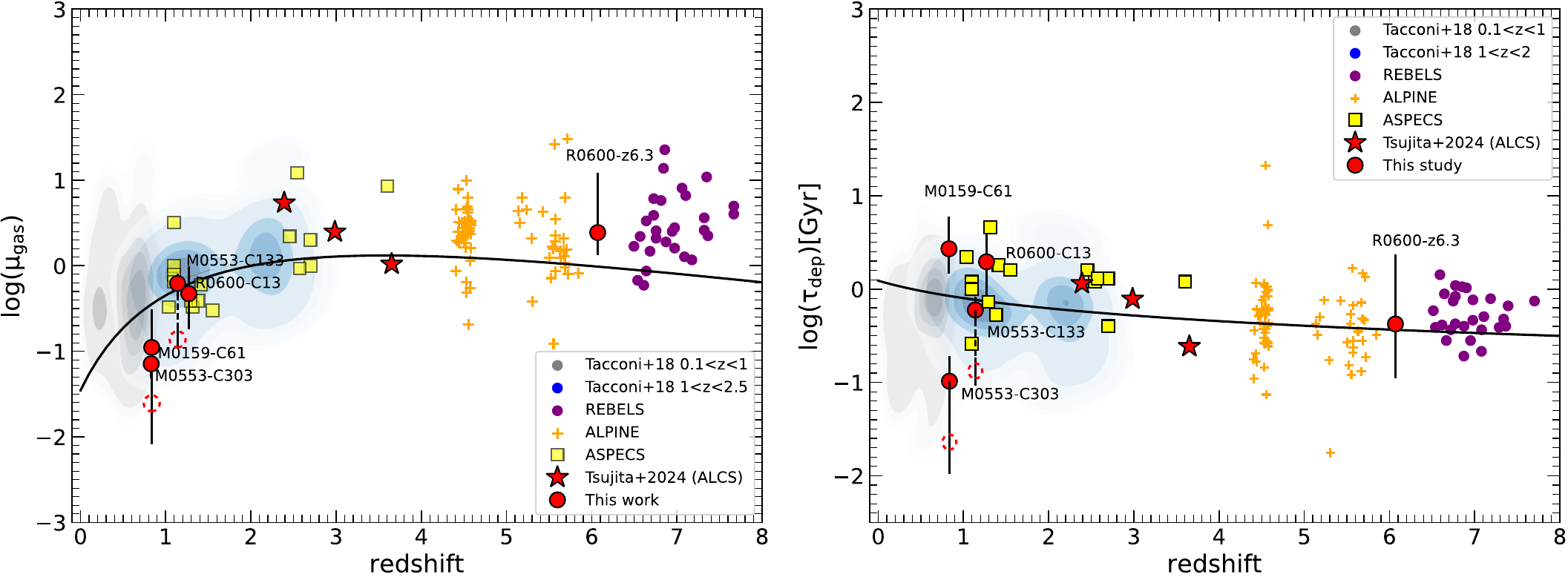}
\caption{
Redshift evolution of molecular gas fraction (left) and gas depletion time scale (right).
ALCS sources (including sources of \citealt{2024arXiv240609890T}) are plotted with red points.  
For comparison, ALPINE and REBELS sources are also plotted with yellow crosses and purple circles, respectively. For the PHIBSS1/2 CO sources \citep{Tacconi_2013,2018ApJ...853..179T}, galaxies at $ 0.1< z < 1$ are shown in gray contours, while those at $1 < z < 2.5$ are displayed in light blue contours.
The solid line represents the scaling relation of star-forming galaxies derived by \cite{2018ApJ...853..179T} where only the redshift dependence is considered, ignoring the contributions from stellar mass and the offset from the main sequence line ($\delta$MS). For sources with multiple possible line identifications, M0553–C303 and R0600–C13 are plotted assuming CO(4–3) and CO(5–4), respectively, following the identifications adopted in \cite{2024ApJS..275...36F}.}
\label{fig:comic_evolution}
\end{figure}

\begin{figure}
\epsscale{1}
\plotone{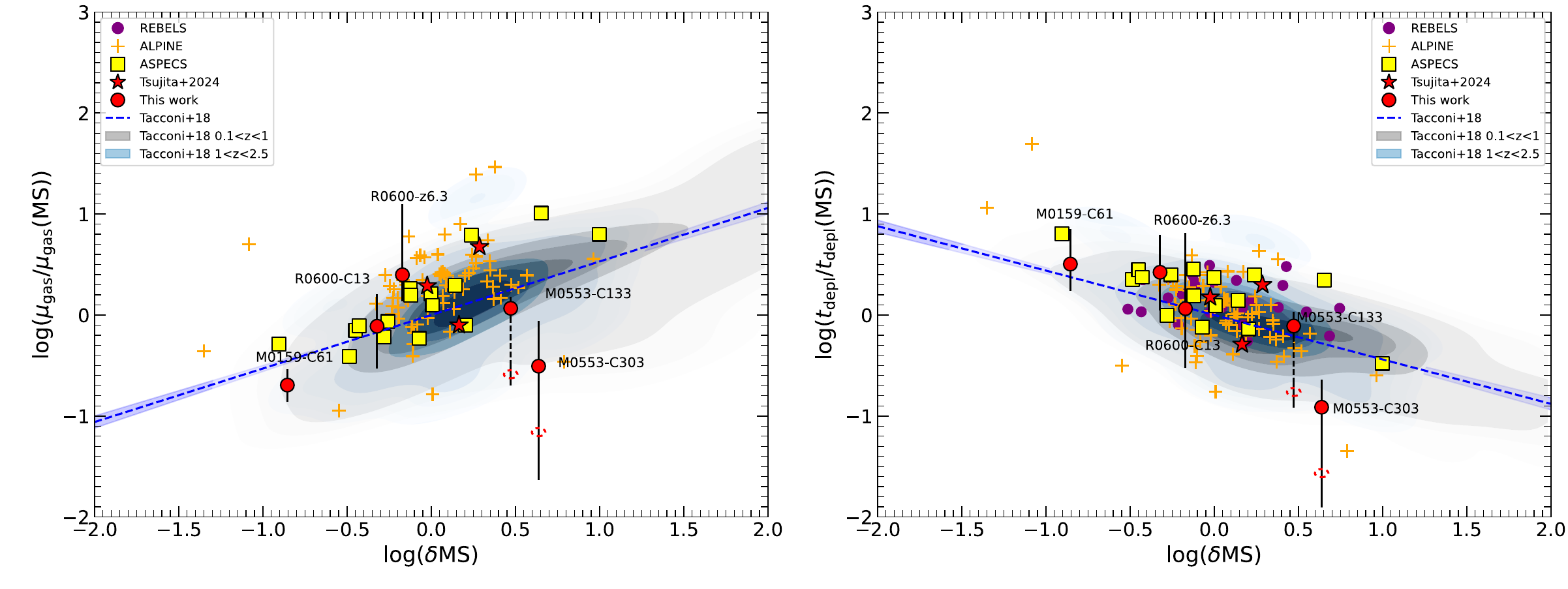}
\caption{
Molecular gas fraction (left) and gas depletion time (right) as a function of the offset from the reference MS line after removal of redshift dependence. 
ALCS sources (including sources of \citealt{2024arXiv240609890T}) are plotted with red points.
For comparison, ALPINE and REBELS sources are also plotted with yellow crosses and purple circles, respectively. For the PHIBSS1/2 CO sources \citep{Tacconi_2013,2018ApJ...853..179T}, galaxies at $0.1< z < 1$ are shown in gray contours, while those at $1 < z < 2.5$ are displayed in light blue contours.
The dashed line shows the best-fit line for the MS galaxies derived in \cite{2018ApJ...853..179T}. For sources with multiple possible line identifications, M0553–C303 and R0600–C13 are plotted assuming CO(4–3) and CO(5–4), respectively, following the identifications adopted in \cite{2024ApJS..275...36F}.}

\label{fig:ms_deviation}
\end{figure}

Figure \ref{fig:comic_evolution} shows the redshift evolution of gas fraction (left) and gas depletion time scale (right). 
The black solid lines represent the scaling relation of star-forming galaxies at $z < 4$ derived by \cite{2018ApJ...853..179T}. 
ALCS sources (including sources of \citealt{2024arXiv240609890T}) are plotted with red points.
For comparison, we also plotted the results of PHIBSS1/2 \citep{Tacconi_2013,2018ApJ...853..179T}, ASPECS \citep{Aravena_2019}, ALPINE \citep{2020A&A...643A...5D}, and REBELS \citep{2024A&A...682A..24A}. 
The black lines in Figure~\ref{fig:comic_evolution} account for the redshift dependence only, and do not incorporate offsets from the main sequence (MS) or variations in specific star formation rate (sSFR = SFR/$M_*$). 
Here, we define the gas fraction as $\mu = M_{\rm gas}/M_\ast$, and the main-sequence offset as $\delta {\rm MS} = \log_{10}({\rm sSFR}/{\rm sSFR}_{\rm MS})$.
It is known that in star-forming galaxies, both the gas fraction and depletion time scale depend not only on redshift but also on sSFR and the offset from the MS. Based on this, Figure~\ref{fig:ms_deviation} shows the normalized gas fraction and depletion time scale as a function of the offset from the MS ($\delta$MS).
The ALCS sources, spanning redshifts from $z \sim 1$ to $z = 6.071$, are broadly consistent with the established scaling relations between molecular gas fraction and depletion time. 
Among them, most $z \sim 1$ emitters lie within the observed scatter, and the [C\,{\sc ii}] emitter R0600-z6.3 at $z = 6.071$ also aligns with the distributions of ALPINE and REBELS galaxies. 
This indicates that the star formation efficiency and gas fraction of typical star-forming galaxies may not deviate significantly from the canonical scaling relations even at early cosmic epochs. 

However, several sources of systematic uncertainty remain. 
Uncertainties in line identification, assumed line luminosity ratios (e.g., from high–$J$ CO to CO(1–0)), and the adopted CO–to–H$_2$ conversion factors ($\alpha_\mathrm{CO}$) can cause variations in the inferred molecular gas masses by up to an order of magnitude. 
In addition, stellar masses and SFRs may be affected by SED–fitting degeneracies and potential AGN contamination. 
Indeed, some $z \sim 1$ sources lie slightly below the expected depletion–time relation, likely reflecting elevated SFRs. 
Among them, M0553-C190 (a counter–image of M0553-C133) shows a high CH/CO abundance ratio (see Section~\ref{sec:ch}), characteristic of X–ray–dominated regions (XDRs) and suggestive of possible AGN activity. 
To assess whether such AGN contributions systematically bias the derived gas properties, follow-up spectroscopic observations—such as optical emission-line diagnostics using the BPT diagram—will be necessary. 

These uncertainties imply that, although the scaling relations appear to hold from $z \sim 1$ to $z \sim 6$, their intrinsic scatter may be larger than previously assumed in \cite{2018ApJ...853..179T}. 
Future blind (sub)millimeter line surveys will be essential to improve the statistical robustness of these conclusions.

\subsection{Serendipitous Detection of CH Emission} \label{sec:ch}

\begin{figure}
\epsscale{1}
\plotone{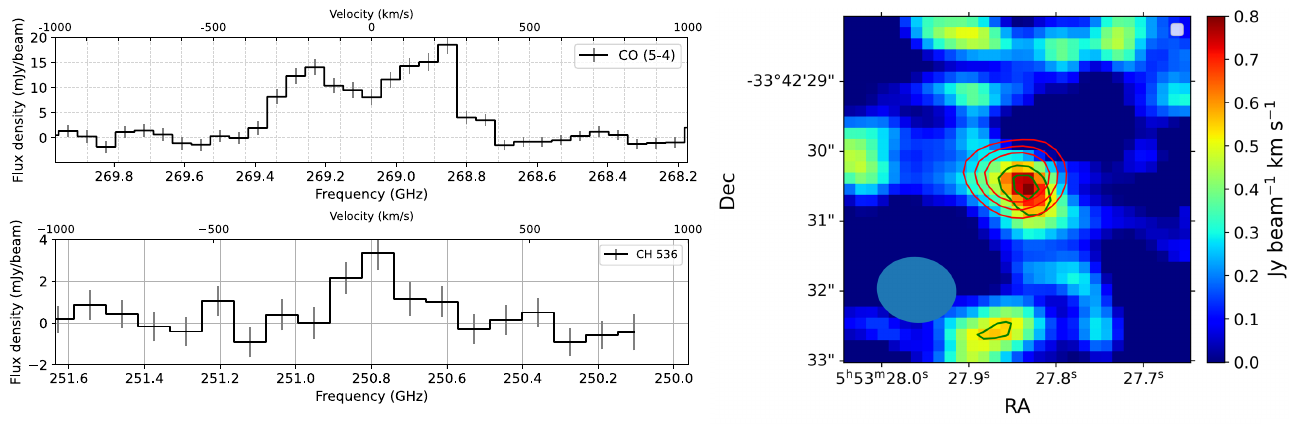}
\caption{Left panel: Line profile of the CO emission (top) and the CH emission (bottom) detected in M0553-C190. Right panel: Velocity-integrated intensity map of the CH emission ($5'' \times 5''$). 
The beam size is shown in the bottom left of the image. Contours are drawn at levels of $3\sigma$ and $4\sigma$, where $\sigma$ is the rms noise of the integrated map. Green contours indicate CH emission, while red contours show CO emission.
}
\label{fig:ch}
\end{figure}

We report a tentative detection of the CH ($N = 1$, $J = 3/2 \rightarrow 1/2$) emission line toward M0553-C190 at $z = 1.14$. 
The spectroscopic redshift was confirmed from optical/NIR spectroscopy \citep{2017MNRAS.471.3305E}, and we also detected the CO(5--4) emission line in this study, showing a clear double-peaked profile (Figures \ref{fig:spe1} and \ref{fig:ch}). 
This CH transition consists of a $\Lambda$-doublet with multiple hyperfine components. The three strongest lines are centered at rest-frame frequencies of 536.7611\,GHz ($F = 2^- \rightarrow 1^+$), 536.7819\,GHz ($F = 1^- \rightarrow 1^+$), and 536.7957\,GHz ($F = 1^- \rightarrow 0^+$), as listed in the Cologne Database for Molecular Spectroscopy (CDMS; \citealt{2001A&A...370L..49M, 2005JMoSt.742..215M}) and the Jet
Propulsion Laboratory (JPL) Submillimeter, Millimeter, and Microwave Spectral Line Catalog (\citealt{1998JQSRT..60..883P}). These components are typically blended into a single broad feature in extragalactic observations due to large velocity dispersions.

The line profile and velocity-integrated intensity map of the CH emission are shown in Figure \ref{fig:ch}.
The line is detected at a significance level of \(4.4\sigma\).
Due to the modest signal-to-noise ratio, it is not possible to distinguish whether the CH line profile corresponds to a single- or double-peaked structure. As discussed in \cite{2014ApJ...785..149S}, an HCO$^{+}$(6--5) line might be expected near the CH 536 GHz transition. In this source, it would be observed at 250.02 GHz, but it is not within our frequency coverage. 
This detection represents the first report of CH emission from a galaxy identified in a blind line survey at cosmological distances, whereas CH emission from an individual high-redshift source has previously been detected toward the quasar APM~08279+5255 at $z=3.91$ \citep{2023AA...680A..95Y}.
\citet{2014ApJ...785..149S} reported CH emission in a stacked ALMA 3 mm spectrum of 22 dusty star-forming galaxies spanning \(z = 2\)–5.7.
Their detection suggested that CH is not uncommon in ISM in high-redshift galaxies, but typically too faint to be observed in individual sources. Based on chemical modeling and comparisons with the Milky Way ISM, \citet{2023ApJ...948...44R} and earlier studies (e.g., \citealt{2014ApJ...785..149S}) suggest that CH likely arises in photon-dominated regions (PDRs) at moderate densities (\(n \sim 10^3\ \mathrm{cm}^{-3}\)), where UV radiation drives the chemistry. No strong evidence for X-ray-dominated regions (XDRs) or AGN-related processes was found in their sample.

In the local universe, CH emission was observed in both starburst- and AGN-dominated galaxies using Herschel \citep{2014ApJ...788..147R}. The CH/CO abundance ratio was found to be significantly higher in AGN hosts such as NGC 1068 than in purely star-forming systems, and XDR models were shown to reproduce these enhanced CH abundances. These results imply that CH can trace chemically distinct gas phases associated with either UV-irradiated diffuse gas (PDRs) or X-ray-irradiated dense gas (XDRs). 
We estimate the column density of CH by using Equation \ref{eq:CH_column_density}. 
Here we assume an excitation temperature of 37.5~K and a background temperature of 5.84~K, corresponding to the CMB temperature at redshift \( z = 1.14 \). The excitation temperature is chosen to be close to the value reported for CH in Arp~220 (\( T_{\mathrm{ex}} \sim 41~\mathrm{K} \); \citealt{2014ApJ...788..147R}), which was directly estimated in Arp~220 and also assumed for other galaxies in that study due to a lack of direct measurements. This serves as an approximate reference rather than a precise value, given that Arp~220 is an extreme starburst galaxy with unique properties (e.g., high dust optical depth and a heavily obscured nucleus). At this excitation temperature, we adopt a partition function value of 18.8471 from the CDMS database.
An integrated brightness temperature was calculated as $\int T_\mathrm{b}(v)\, dv \approx 15\ \mathrm{K\ km\ s^{-1}}$ after lensing correction. The derived column density is $N_\mathrm{CH} \approx 2.0 \times 10^{14}\ \mathrm{cm}^{-2}.$
To estimate the CO column density, we extrapolate the CO($1$--$0$) brightness temperature from the observed CO($5$--$4$) line using the empirical ratio \citep{daddi2015}. Given \( \int T_\mathrm{b}(5\text{--}4)\, dv \approx 18\ \mathrm{K\ km\ s^{-1}} \), we infer \( \int T_\mathrm{b}(1\text{--}0)\, dv \approx 80\ \mathrm{K\ km\ s^{-1}} \). Using the Galactic conversion factor \( N(\mathrm{H}_2) = 2 \times 10^{20} \int T_\mathrm{b}(1\text{--}0)\, dv\ \mathrm{cm^{-2}} \), we obtain \( N(\mathrm{H}_2) \approx 1.6 \times 10^{22}\ \mathrm{cm^{-2}} \).
Assuming a CO abundance relative to H$_2$ of $X(\mathrm{CO}) =  10^{-4}$ as a canonical value for Galactic disk clouds (\citealt{2008ApJ...679..481P}, and
references therein), the CO column density is estimated to be $N(\mathrm{CO}) \approx 1.6 \times 10^{18}\ \mathrm{cm}^{-2}.$
Combining this with our earlier estimate for CH, we find the CH-to-CO column density ratio to be $N_\mathrm{CH}/N_\mathrm{CO} \approx 1.3 \times 10^{-4}$. 


\begin{table}[ht]
\centering
\caption{CH-to-CO column density ratio in different galaxy types}
\label{tab:CHCO}
\begin{tabular}{lccc}
\hline
Object & $N_\mathrm{CH}/N_\mathrm{CO}$ & Reference & Galaxy type \\
\hline
Arp 220        & $4.5\times10^{-5}$          & \cite{2014ApJ...788..147R} & ULIRG \\ 
NGC 1068       & $1\times10^{-4}$            & \cite{2012ApJ...758..108S, 2014ApJ...788..147R} & AGN (Seyfert 2) \\
NGC 253        & $2\times10^{-5}$            & \cite{2014ApJ...788..147R} & Starburst \\
M82            & $2\times10^{-5}$            & \cite{2014ApJ...788..147R} & Starburst \\
APM~08279+5255 & $8\times10^{-3}$            & \cite{2023AA...680A..95Y} & Lensed QSO ($z=3.91$) \\
M0553-C190     & $(1.3\pm 0.3)\times10^{-4}$ & This work & Lensed star-forming galaxy ($z=1.14$) \\
\hline
\end{tabular}
\end{table}

Table \ref{tab:CHCO} shows the CH-to-CO column density ratio, $N_\mathrm{CH}/N_\mathrm{CO}$. We refer to previous studies (e.g., Arp 220: starburst/AGN composite; NGC 1068: Seyfert 2 AGN; M82 and NGC 253: starburst galaxies without AGN; e.g., \citealt{2014ApJ...788..147R}) and plot the CH/CO ratios for each galaxy.
In starburst galaxies without AGN, the $N_\mathrm{CH}/N_\mathrm{CO}$ ratios are relatively low, around $10^{-5}$. In contrast, galaxies hosting AGN exhibit higher ratios, typically on the order of $10^{-4}$ — about one order of magnitude higher than those without AGN. Intermediate systems show values between $10^{-5}$ and $10^{-4}$, suggesting a gradual enhancement of CH abundance relative to CO depending on the AGN contribution.
Hard X-ray photons from an AGN are capable of penetrating much larger column densities than UV photons originating from PDRs, thereby enabling the reactions 
to proceed deeper into molecular clouds, enhancing CH abundance (\citealt{1973ApL....15...79B}; \citealt{2014ApJ...788..147R}). 
Furthermore, in XDR regions, radiative association reactions between neutral carbon and hydrogen ($\mathrm{C}+\mathrm{H} \rightarrow \mathrm{CH}+h\nu$), as well as neutral-neutral reactions ($\mathrm{C}+\mathrm{H}_{2} \rightarrow \mathrm{CH}+\mathrm{H}$), become dominant, significantly enhancing CH production \citep{2014ApJ...788..147R}.
Therefore, $X(\mathrm{CH})/X(\mathrm{CO})$ may serve as a useful diagnostic to distinguish between PDR- and XDR-dominated environments, that is, whether a galaxy is purely star-forming or hosts a significant AGN contribution. We also note that CH emission has been reported toward the lensed quasar APM~08279+5255 at $z=3.91$ \citep{2023AA...680A..95Y}, which shows an extremely high CH-to-CO ratio of $\sim 8\times10^{-3}$. This value is significantly higher than those found in local starburst galaxies and even Seyfert AGN, suggesting that CH can be strongly enhanced in AGN-dominated environments and that elevated CH/CO ratios may serve as a diagnostic of X-ray irradiated gas.
As shown in Table \ref{tab:CHCO}, the high $N_\mathrm{CH}/N_\mathrm{CO}$ ratio observed in the high-redshift galaxy M0553-C190 may suggest the presence of an AGN contribution. Nevertheless, there remains uncertainty due to the assumed CO excitation conditions. Constraining the CO SLED of this source through multi-wavelength follow-up observations is crucial to obtain definitive evidence for AGN activity. Considering that CH was detected even in a blind search, it may serve as a promising probe for AGN diagnostics at high redshift in future studies. Further details will be presented in our forthcoming paper.


\subsection{Nature of Unidentified Line Emitters Candidates Without Counterparts}\label{sec:ud}
Among the 16 emission line candidates detected in our blind search, four sources (AC0102-EL01, AS1063-EL01, M1206-EL01, and A2163-EL01) exhibit no detectable counterparts at any other wavelengths, including deep HST, IRAC, and Herschel (Figure~\ref{fig:img}). The lack of multi-wavelength identification prevents robust redshift determination and line identification. 
None of the unidentified lines were found to coincide with emission lines at the cluster redshift.
In this section, we discuss the possible nature of these sources using gravitational lensing models and evaluate the feasibility of various redshift scenarios based on the lensing geometry and magnification.

\begin{description}
\item[AC0102-EL01] \mbox{}\\
AC0102-EL01 is located on the outskirts of the ACTCLJ0102--4915 cluster. 
Even assuming a high redshift of $z = 6$, where gravitational magnification is typically expected to be large, the predicted magnification remains modest ($\mu \sim 1.2$), implying that the observed line flux is close to its intrinsic value. 
This field lies outside the available James Webb Space Telescope (JWST) imaging footprint \citep{2023ApJ...952...81F}, limiting further constraints at present.

\item[AS1063-EL01] \mbox{}\\
AS1063-EL01 is located near the core of the Abell S1063 cluster. For hypothetical redshifts of $z = 1$, 2, and 6, GLAFIC lensing models \citep{2018ApJ...855....4K} predict three multiple images with similar magnifications for $z=2$ and 6. However, no significant line emission is detected at the predicted positions of the counter images. 
Despite the availability of deep JWST imaging \citep{2025ApJ...983L..22K}, no NIR counterpart is found at the source position in the NIRCam F115W and F444W filters.

\item[M1206-EL01] \mbox{}\\
M1206-EL01 is a particularly intriguing source located near the critical curve of the MACS1206 cluster. For illustration, we assume $z = 6$, since the lensing magnification increases with source redshift up to $z \sim 1$--2 and then saturates at higher redshifts; thus, this choice would also be approximately valid if the line corresponds to [C\,{\sc ii}] (see Figure \ref{fig:redshift}).
The GLAFIC lens model \citep{2020MNRAS.496.2591O} predicts extremely high magnifications ($\mu \gtrsim 35$) and two multiple images with $\mu = 37$ and $-36$
straddling the critical curve. Considering uncertainties in the mass model and source redshift, the two highly magnified images may have an angular separation comparable to or smaller than the observational resolution, thus appearing as a single blended source, and have even higher magnifications. In this case, the total magnification corresponds to the sum of the absolute magnifications of the two multiple images, potentially reaching several hundred if the source is intrinsically sufficiently compact. Additionally, a third image with lower magnification ($\mu \approx 3.1$) is predicted at a more distant position, which is consistent with no detection in our observation due to its lower magnification. Consequently, M1206-EL01 can be interpreted as a multiply imaged system, with only the blended, highly magnified component being observed. 
The strong magnification likely enables the detection of the line emission, while the absence of continuum in HST imaging suggests that the source might be deeply dust-obscured and invisible in the UV–optical regime.

\item[A2163-EL01] \mbox{}\\
A2163-EL01 is located in a relatively outer part of the Abell~2163 cluster. Lensing models predict modest magnifications ($\mu \sim 2.1$--$2.6$) for $z = 1$--6, and only a single image is expected. The moderate magnification implies that the observed flux is not significantly boosted.

\end{description}

Although the current data do not allow a definitive identification of these four sources, gravitational lensing models provide valuable constraints. Among them, M1206-EL01 stands out as a potentially interesting candidate for a highly magnified, optically invisible galaxy at high redshift. Overall, the absence of multi-wavelength counterparts suggests that these sources are either intrinsically low-luminosity or heavily dust-obscured galaxies. On the other hand, instead of being noise, they could also represent unlensed galaxies at the cluster redshift or lower redshifts. Future deep NIR or submillimeter observations will be crucial for testing these scenarios and further exploring the nature of these optically dark line emitters, which may represent a hidden population of faint or dust-enshrouded galaxies that have been uncovered by blind millimeter surveys.

\section{Conclusions}\label{sec:summary}
We conducted a blind search for line-emitting galaxies using data from the ALMA Lensing Cluster Survey (ALCS), a Cycle 6 ALMA Large Program. Our primary findings are as follows:

\begin{enumerate}
\item We robustly detected seven line emitters, including one [C\,{\sc ii}] emitter at $z = 6.071$, four CO emitters at $z = 0.8$–$1.1$, and two additional sources whose lines are consistent with either CO or [C\,{\sc i}] within their photometric redshift ranges. Additionally, we identified four unidentified emission-line candidates lacking counterparts at other wavelengths.

\item Thanks to gravitational lensing, our sample probes molecular gas masses $\sim1$ dex below the lower bound of previous ALMA deep surveys, extending the exploration toward intrinsically fainter, gas-poor galaxies.
While the $z\sim1$ sources are broadly consistent with established scaling relations between gas fraction and depletion time, the uncertainties in line identification and conversion factors prevent us from drawing firm conclusions about systematic offsets.
In particular, adopting line ratios or CO--to--H$_2$ conversion factors derived for starburst galaxies could overestimate the gas masses by up to an order of magnitude, potentially placing some sources as outliers from the canonical relations.
This highlights that the widely used scaling relations may possess larger intrinsic scatter than previously assumed.

\item The gas fraction and depletion timescale of our [C\,{\sc ii}] emitter at $z = 6.071$ align with those derived from galaxies selected in the UV, optical, and near-infrared, implying no significant selection bias between UV-selected and (sub)millimeter line-selected galaxies at high redshift.

\item We report a tentative detection of CH ($N=1$, $J=3/2 \rightarrow 1/2$) emission at $z = 1.14$ in the gravitationally lensed galaxy M0553-C190, representing the first CH detection from a galaxy identified in a blind line survey at cosmological distances.

\item We also identified four emission-line candidates without any counterparts in deep optical-to-infrared images. Although their nature remains uncertain, lensing models suggest they could be heavily dust-obscured or intrinsically faint high-redshift galaxies. Future deep NIR or submillimeter follow-up observations will be essential to clarify their nature and reveal this potentially hidden galaxy population.

\end{enumerate}

Our findings underscore the importance of gravitational lensing in revealing faint galaxy populations and offer new insights into the molecular gas properties of galaxies across cosmic time.

\begin{acknowledgments}
We would like to acknowledge the referee for helpful comments and suggestions. We are grateful to Manuel Aravena for generously providing the REBELS data, which greatly contributed to this work.
We thank Kazuhiro Shimasaku and Yuichi Harikane for fruitful
discussions.
This work was supported by JST SPRING, Grant Number JPMJSP2108.
KK acknowledges the support by JSPS KAKENHI Grant Numbers JP22H04939, 23K20035, and 24H00004.
K.N. was supported by the International Graduate Program for Excellence in Earth-Space Science (IGPEES), the University of Tokyo.
K.N. was supported by the ALMA Japan Research Grant of NAOJ ALMA Project, NAOJ-ALMA-367.
JGL acknowledges support from ANID BASAL project FB210003 and Programa de Inserción Académica 2024, Vicerrectoría Académica y Prorrectoría, Pontificia Universidad Católica de Chile.
MO acknowledges the support by JSPS KAKENHI Grant Numbers JP22K21349, JP25H00662.
This paper uses the ALMA data: ADS/JAO. ALMA$\#$ 2013.1.00999.S,  ADS/JAO. ALMA$\#$2015.1.01425.S and ADS/JAO. ALMA $\#$2018.1.00035.L. 
ALMA is a partnership of ESO (representing its member states), NSF (USA), and NINS (Japan), together with NRC (Canada), MOST and ASIAA (Taiwan), and KASI (Republic of Korea), in cooperation with the Republic of Chile. The Joint ALMA Observatory is operated by ESO, AUI/NRAO and NAOJ. The National Radio Astronomy Observatory is a facility of the National Science
Foundation operated under cooperative agreement by Associated Universities, Inc.
Data analysis was conducted using the Multiwavelength Data Analysis System, operated by the Astronomy Data Center (ADC) at the National Astronomical Observatory of Japan. 
This work is based on observations and archival data acquired using the
Spitzer Space Telescope, which is operated by the Jet
Propulsion Laboratory, California Institute of Technology,
under a contract with NASA, along with archival data from the
NASA/ESA Hubble Space Telescope. This research also made use of the NASA/IPAC Infrared Science Archive (IRSA),
which is operated by the Jet Propulsion Laboratory, California
Institute of Technology, under contract with the National
Aeronautics and Space Administration.

\end{acknowledgments}
%

\vspace{5mm}
\software{Astropy \citep{2022ApJ...935..167A}, Glafic \citep{2010PASJ...62.1017O}, CASA \citep{2007ASPC..376..127M,cite-key}, \texttt{easy} \citep{2008ApJ...686.1503B}, \texttt{MAGPHYS} \citep{2015ApJ...806..110D}}

\appendix
\section{Overview of the detected line-emitting galaxies}\label{sec:indiv}

\subsubsection*{R0600-z6.3}
R0600-z6.3 was detected toward RXCJ0600.1$-$2007. The photometric redshift estimated from the \texttt{easy} SED fitting is \( z_{\mathrm{phot}} = 6.09^{+0.09}_{-0.08} \), and the observed line frequency is \( \nu_{\mathrm{obs}} = 268.747 \)~GHz. Based on these values, the line can be uniquely identified as [C\,{\sc ii}] from among the possible line candidates listed in Table~\ref{table:archive-1a}.
The detection and physical properties of this source were reported in  \cite{Fujimoto_2021}, \cite{2021MNRAS.505.4838L}, and \cite{2024arXiv240218543F}. R0600-z6.3 has two additional lensed images identified in the previous studies. One of them, designated as R0600-C164 in \cite{2024ApJS..275...36F}, was detected in ALMA 1.2 mm continuum.

\subsubsection*{M0553-C190/M0553-C249/M0553-C133}
M0553-C190, M0553-C249, and M0553-C133 were detected toward MACSJ0553.4$-$3342 and correspond to multiple lensed images of the same source. The emission line was observed at 269.070 GHz and is uniquely identified as CO(5--4) at $z=1.14$, consistent with the spectroscopic redshift reported by \citet{2017MNRAS.471.3305E}. These sources were also reported by \citet{2024ApJS..275...36F}.

\subsubsection*{M0553-C303 }
M0553-C303 was detected toward MACSJ0553.4-3342. The observed frequency was 250.646 GHz. Since the source was located outside the HST footprint, it was not included in the \texttt{easy} catalog. 
The photometric redshift was estimated to be $z_\mathrm{phot} = 0.98_{-0.39}^{+0.39}$ from the FIR SED fitting by \cite{2022ApJ...932...77S}, which is consistent with the interpretation of the line as CO(4--3) at $z = 0.836$, though this does not necessarily represent a unique identification. Possible line identifications include CO(4--3) ($z = 0.836$), 
[C\,{\sc i}](1--0) ($z = 0.96$), and CO(5--4) ($z = 1.30$), with relative likelihoods of approximately 19\%, 21\%, and 14\%, respectively. \cite{2024ApJS..275...36F} also reported this source and adopted CO(4--3) as the most plausible identification; however, in this paper we consider multiple possible line identifications for this and other ambiguous sources. Follow-up observations will be required to confirm the true origin of the line.

\subsubsection*{M0159-C61}
M0159-C61 was detected toward MACSJ0159.8$-$0849. The photometric redshift estimated from the \texttt{easy} SED fitting is \( z_{\mathrm{phot}} = 0.83^{+0.10}_{-0.04} \), and the observed line frequency is \( \nu_{\mathrm{obs}} = 251.862\)~GHz. Based on these values, the line is identified as CO(4--3) at \( z = 0.84 \). 

\subsubsection*{R0600-C13}
R0600-C13 was detected toward RXCJ0600.1-2007. As the source falls outside of the HST/WFC3 images, it was not included in the catalog identified by Lineseeker.
In \cite{2024ApJS..275...36F}, a template fitting analysis incorporating multi-wavelength data yields a photometric redshift of $1.00_{-0.05}^{+0.05}$, consistent with the result obtained independently through SED fitting using \texttt{MAGPHYS} ($1.23_{-0.45}^{+0.40}$). Possible line identifications include CO(5--4) at $z = 1.27$, [C\,\textsc{i}](1--0) at $z = 0.94$, and CO(4--3) at $z = 0.82$, with relative likelihoods of approximately 18.7\%, 15.4\%, and 12.5\%, respectively. Based on \texttt{MAGPHYS} results, we regard CO(5--4) as the most plausible identification, in agreement with the interpretation presented by \cite{2024ApJS..275...36F}. However, as in the case of M0553--C303, we consider multiple possible line identifications for this and other ambiguous sources. Future follow-up observations, which may detect additional CO transitions and determine a spectroscopic redshift, will be required to confirm the true origin of the line.

\section{Deriving column density of CH}
Under the assumption of optically thin emission ($\tau \ll 1$), the column density of CH is given by 

\begin{equation}
N_\mathrm{CH} = \frac{8 \pi \nu^3}{c^3 A_{ul}} \cdot \frac{Q(T_\mathrm{ex})}{g_u} \exp\left(\frac{E_u}{T_\mathrm{ex}}\right) \cdot \frac{1}{J_\nu(T_\mathrm{ex}) - J_\nu(T_\mathrm{bg})} \cdot \int T_\mathrm{b}(v)\, dv,
\label{eq:CH_column_density}
\end{equation}
where $\nu$ is the rest frequency, $A_{ul}$ is the Einstein A-coefficient, $Q(T_\mathrm{ex})$ is the partition function, $g_u$ is the degeneracy of the upper level, $E_u$ is the upper level energy, and $J_\nu(T)$ is the Planck function in temperature units \citep{2015PASP..127..266M}.

\bibliography{sample631}{}

@article{Solomon_1997,
	author = {P. M. Solomon and D. Downes and S. J. E. Radford and J. W. Barrett},
    title = {The Molecular Interstellar Medium in Ultraluminous Infrared Galaxies},
	doi = {10.1086/303765},
	journal = {The Astrophysical Journal},
	month = {mar},
	number = {1},
	pages = {144},
	url = {https://dx.doi.org/10.1086/303765},
	volume = {478},
	year = {1997}}

@ARTICLE{2018MNRAS.481.1976Z,
       author = {{Zanella}, A. and {Daddi}, E. and {Magdis}, G. and {Diaz Santos}, T. and {Cormier}, D. and {Liu}, D. and {Cibinel}, A. and {Gobat}, R. and {Dickinson}, M. and {Sargent}, M. and {Popping}, G. and {Madden}, S.~C. and {Bethermin}, M. and {Hughes}, T.~M. and {Valentino}, F. and {Rujopakarn}, W. and {Pannella}, M. and {Bournaud}, F. and {Walter}, F. and {Wang}, T. and {Elbaz}, D. and {Coogan}, R.~T.},
        title = "{The [C II] emission as a molecular gas mass tracer in galaxies at low and high redshifts}",
      journal = {\mnras},
         year = 2018,
        month = dec,
       volume = {481},
       number = {2},
        pages = {1976-1999},
          doi = {10.1093/mnras/sty2394},
archivePrefix = {arXiv},
       eprint = {1808.10331},
 primaryClass = {astro-ph.GA},
       adsurl = {https://ui.adsabs.harvard.edu/abs/2018MNRAS.481.1976Z}
}

@article{Postman_2012,
doi = {10.1088/0067-0049/199/2/25},
url = {https://dx.doi.org/10.1088/0067-0049/199/2/25},
year = {2012},
month = {mar},
publisher = {The American Astronomical Society},
volume = {199},
number = {2},
pages = {25},
author = {Marc Postman and Dan Coe and Narciso Benítez and Larry Bradley and Tom Broadhurst and Megan Donahue and Holland Ford and Or Graur and Genevieve Graves and Stephanie Jouvel and Anton Koekemoer and Doron Lemze and Elinor Medezinski and Alberto Molino and Leonidas Moustakas and Sara Ogaz and Adam Riess and Steve Rodney and Piero Rosati and Keiichi Umetsu and Wei Zheng and Adi Zitrin and Matthias Bartelmann and Rychard Bouwens and Nicole Czakon and Sunil Golwala and Ole Host and Leopoldo Infante and Saurabh Jha and Yolanda Jimenez-Teja and Daniel Kelson and Ofer Lahav and Ruth Lazkoz and Dani Maoz and Curtis McCully and Peter Melchior and Massimo Meneghetti and Julian Merten and John Moustakas and Mario Nonino and Brandon Patel and Enikö Regös and Jack Sayers and Stella Seitz and Arjen Van der Wel},
title = {THE CLUSTER LENSING AND SUPERNOVA SURVEY WITH HUBBLE: AN OVERVIEW},
journal = {The Astrophysical Journal Supplement Series}
}

@ARTICLE{2017ApJ...837...97L,
       author = {{Lotz}, J.~M. and {Koekemoer}, A. and {Coe}, D. and {Grogin}, N. and {Capak}, P. and {Mack}, J. and {Anderson}, J. and {Avila}, R. and {Barker}, E.~A. and {Borncamp}, D. and {Brammer}, G. and {Durbin}, M. and {Gunning}, H. and {Hilbert}, B. and {Jenkner}, H. and {Khandrika}, H. and {Levay}, Z. and {Lucas}, R.~A. and {MacKenty}, J. and {Ogaz}, S. and {Porterfield}, B. and {Reid}, N. and {Robberto}, M. and {Royle}, P. and {Smith}, L.~J. and {Storrie-Lombardi}, L.~J. and {Sunnquist}, B. and {Surace}, J. and {Taylor}, D.~C. and {Williams}, R. and {Bullock}, J. and {Dickinson}, M. and {Finkelstein}, S. and {Natarajan}, P. and {Richard}, J. and {Robertson}, B. and {Tumlinson}, J. and {Zitrin}, A. and {Flanagan}, K. and {Sembach}, K. and {Soifer}, B.~T. and {Mountain}, M.},
        title = "{The Frontier Fields: Survey Design and Initial Results}",
      journal = {\apj},
         year = 2017,
        month = mar,
       volume = {837},
       number = {1},
          eid = {97},
        pages = {97},
          doi = {10.3847/1538-4357/837/1/97},
archivePrefix = {arXiv},
       eprint = {1605.06567},
 primaryClass = {astro-ph.GA},
       adsurl = {https://ui.adsabs.harvard.edu/abs/2017ApJ...837...97L}
}

@ARTICLE{2019ApJ...884...85C,
       author = {{Coe}, Dan and {Salmon}, Brett and {Brada{\v{c}}}, Maru{\v{s}}a and {Bradley}, Larry D. and {Sharon}, Keren and {Zitrin}, Adi and {Acebron}, Ana and {Cerny}, Catherine and {Cibirka}, Nath{\'a}lia and {Strait}, Victoria and {Paterno-Mahler}, Rachel and {Mahler}, Guillaume and {Avila}, Roberto J. and {Ogaz}, Sara and {Huang}, Kuang-Han and {Pelliccia}, Debora and {Stark}, Daniel P. and {Mainali}, Ramesh and {Oesch}, Pascal A. and {Trenti}, Michele and {Carrasco}, Daniela and {Dawson}, William A. and {Rodney}, Steven A. and {Strolger}, Louis-Gregory and {Riess}, Adam G. and {Jones}, Christine and {Frye}, Brenda L. and {Czakon}, Nicole G. and {Umetsu}, Keiichi and {Vulcani}, Benedetta and {Graur}, Or and {Jha}, Saurabh W. and {Graham}, Melissa L. and {Molino}, Alberto and {Nonino}, Mario and {Hjorth}, Jens and {Selsing}, Jonatan and {Christensen}, Lise and {Kikuchihara}, Shotaro and {Ouchi}, Masami and {Oguri}, Masamune and {Welch}, Brian and {Lemaux}, Brian C. and {Andrade-Santos}, Felipe and {Hoag}, Austin T. and {Johnson}, Traci L. and {Peterson}, Avery and {Past}, Matthew and {Fox}, Carter and {Agulli}, Irene and {Livermore}, Rachael and {Ryan}, Russell E. and {Lam}, Daniel and {Sendra-Server}, Irene and {Toft}, Sune and {Lovisari}, Lorenzo and {Su}, Yuanyuan},
        title = "{RELICS: Reionization Lensing Cluster Survey}",
      journal = {\apj},
         year = 2019,
        month = oct,
       volume = {884},
       number = {1},
          eid = {85},
        pages = {85},
          doi = {10.3847/1538-4357/ab412b},
archivePrefix = {arXiv},
       eprint = {1903.02002},
 primaryClass = {astro-ph.GA},
       adsurl = {https://ui.adsabs.harvard.edu/abs/2019ApJ...884...85C}
}

@article{ refId0,
	author = {{González-López, J.} and {Bauer, F. E.} and {Romero-Cañizales, C.} and {Kneissl, R.} and {Villard, E.} and {Carvajal, R.} and {Kim, S.} and {Laporte, N.} and {Anguita, T.} and {Aravena, M.} and {Bouwens, R. J.} and {Bradley, L.} and {Carrasco, M.} and {Demarco, R.} and {Ford, H.} and {Ibar, E.} and {Infante, L.} and {Messias, H.} and {Muñoz Arancibia, A. M.} and {Nagar, N.} and {Padilla, N.} and {Treister, E.} and {Troncoso, P.} and {Zitrin, A.}},
	title = {The ALMA Frontier Fields Survey - I. 1.1 mm continuum detections in Abell 2744, MACS J0416.1-2403   and MACS J1149.5+2223⋆},
	DOI= "10.1051/0004-6361/201628806",
	url= "https://doi.org/10.1051/0004-6361/201628806",
	journal = {A\&A},
	year = 2017,
	volume = 597,
	pages = "A41",
}

@INPROCEEDINGS{2007ASPC..376..127M,
       author = {{McMullin}, J.~P. and {Waters}, B. and {Schiebel}, D. and {Young}, W. and {Golap}, K.},
        title = "{CASA Architecture and Applications}",
    booktitle = {Astronomical Data Analysis Software and Systems XVI},
         year = 2007,
       editor = {{Shaw}, R.~A. and {Hill}, F. and {Bell}, D.~J.},
       series = {Astronomical Society of the Pacific Conference Series},
       volume = {376},
        month = oct,
        pages = {127},
       adsurl = {https://ui.adsabs.harvard.edu/abs/2007ASPC..376..127M}
}

@article{cite-key,
	author = {The CASA Team and Bean, Ben and Bhatnagar, Sanjay and Castro, Sandra and Meyer, Jennifer Donovan and Emonts, Bjorn and Garcia, Enrique and Garwood, Robert and Golap, Kumar and Villalba, Justo Gonzalez and Harris, Pamela and Hayashi, Yohei and Hoskins, Josh and Hsieh, Mingyu and Jagannathan, Preshanth and Kawasaki, Wataru and Keimpema, Aard and Kettenis, Mark and Lopez, Jorge and Marvil, Joshua and Masters, Joseph and McNichols, Andrew and Mehringer, David and Miel, Renaud and Moellenbrock, George and Montesino, Federico and Nakazato, Takeshi and Ott, Juergen and Petry, Dirk and Pokorny, Martin and Raba, Ryan and Rau, Urvashi and Schiebel, Darrell and Schweighart, Neal and Sekhar, Srikrishna and Shimada, Kazuhiko and Small, Des and Steeb, Jan-Willem and Sugimoto, Kanako and Suoranta, Ville and Tsutsumi, Takahiro and van Bemmel, Ilse M. and Verkouter, Marjolein and Wells, Akeem and Xiong, Wei and Szomoru, Arpad and Griffith, Morgan and Glendenning, Brian and Kern, Jeff},
	date = {2022/11/15},
	doi = {10.1088/1538-3873/ac9642},
	isbn = {1538-3873; 0004-6280},
	journal = {Publications of the Astronomical Society of the Pacific},
	number = {1041},
	pages = {114501},
	publisher = {The Astronomical Society of the Pacific},
	title = {CASA, the Common Astronomy Software Applications for Radio Astronomy},
	url = {https://dx.doi.org/10.1088/1538-3873/ac9642},
	volume = {134},
	year = {2022}}

@article{ daddi2015,
	author = {{Daddi,} and {Dannerbauer, H.} and {Liu, D.} and {Aravena, M.} and {Bournaud, F.} and {Walter, F.} and {Riechers, D.} and {Magdis, G.} and {Sargent, M.} and {Béthermin, M.} and {Carilli, C.} and {Cibinel, A.} and {Dickinson, M.} and {Elbaz, D.} and {Gao, Y.} and {Gobat, R.} and {Hodge, J.} and {Krips, M.}},
	title = {CO excitation of normal star-forming galaxies out to z = 1.5 as regulated by the properties of their interstellar medium},
	DOI= "10.1051/0004-6361/201425043",
	url= "https://doi.org/10.1051/0004-6361/201425043",
	journal = {A\&A},
	year = 2015,
	volume = 577,
	pages = "A46",
	month = "",
}

@article{Brammer_2008,
doi = {10.1086/591786},
url = {https://dx.doi.org/10.1086/591786},
year = {2008},
month = {oct},
publisher = {},
volume = {686},
number = {2},
pages = {1503},
author = {Gabriel B. Brammer and Pieter G. van Dokkum and Paolo Coppi},
title = {EAZY: A Fast, Public Photometric Redshift Code},
journal = {The Astrophysical Journal}
}

@article{Kokorev_2022,
doi = {10.3847/1538-4365/ac9909},
url = {https://dx.doi.org/10.3847/1538-4365/ac9909},
year = {2022},
month = {dec},
publisher = {The American Astronomical Society},
volume = {263},
number = {2},
pages = {38},
author = {V. Kokorev and G. Brammer and S. Fujimoto and K. Kohno and G. E. Magdis and F. Valentino and S. Toft and P. Oesch and I. Davidzon and F. E. Bauer and D. Coe and E. Egami and M. Oguri and M. Ouchi and M. Postman and J. Richard and J.-B. Jolly and K. K. Knudsen and F. Sun and J. R. Weaver and Y. Ao and A. J. Baker and L. Bradley and K. I. Caputi and M. Dessauges-Zavadsky and D. Espada and B. Hatsukade and A. M. Koekemoer and A. M. Muñoz Arancibia and K. Shimasaku and H. Umehata and T. Wang and W.-H. Wang},
title = {ALMA Lensing Cluster Survey: Hubble Space Telescope and Spitzer Photometry of 33 Lensed Fields Built with CHArGE},
journal = {The Astrophysical Journal Supplement Series}
}

@article{Fujimoto_2021,
doi = {10.3847/1538-4357/abd7ec},
url = {https://dx.doi.org/10.3847/1538-4357/abd7ec},
year = {2021},
month = {apr},
publisher = {The American Astronomical Society},
volume = {911},
number = {2},
pages = {99},
author = {Seiji Fujimoto and Masamune Oguri and Gabriel Brammer and Yuki Yoshimura and Nicolas Laporte and Jorge González-López and Gabriel B. Caminha and Kotaro Kohno and Adi Zitrin and Johan Richard and Masami Ouchi and Franz E. Bauer and Ian Smail and Bunyo Hatsukade and Yoshiaki Ono and Vasily Kokorev and Hideki Umehata and Daniel Schaerer and Kirsten Knudsen and Fengwu Sun and Georgios Magdis and Francesco Valentino and Yiping Ao and Sune Toft and Miroslava Dessauges-Zavadsky and Kazuhiro Shimasaku and Karina Caputi and Haruka Kusakabe and Kana Morokuma-Matsui and Kikuchihara Shotaro and Eiichi Egami and Minju M. Lee and Timothy Rawle and Daniel Espada},
title = {ALMA Lensing Cluster Survey: Bright [C ii] 158 μm Lines from a Multiply Imaged Sub-L⋆ Galaxy at z = 6.0719},
journal = {The Astrophysical Journal}
}

@article{annurev:/content/journals/10.1146/annurev-astro-081811-125615,
   author = "Madau, Piero and Dickinson, Mark",
   title = "Cosmic Star-Formation History", 
   journal= "Annual Review of Astronomy and Astrophysics",
   year = "2014",
   volume = "52",
   number = "Volume 52, 2014",
   pages = "415-486",
   doi = "https://doi.org/10.1146/annurev-astro-081811-125615",
   url = "https://www.annualreviews.org/content/journals/10.1146/annurev-astro-081811-125615",
   publisher = "Annual Reviews",
   issn = "1545-4282",
   type = "Journal Article",
  }

@article{Bouwens_2015,
doi = {10.1088/0004-637X/803/1/34},
url = {https://dx.doi.org/10.1088/0004-637X/803/1/34},
year = {2015},
month = {apr},
publisher = {The American Astronomical Society},
volume = {803},
number = {1},
pages = {34},
author = {R. J. Bouwens and G. D. Illingworth and P. A. Oesch and M. Trenti and I. Labbé and L. Bradley and M. Carollo and P. G. van Dokkum and V. Gonzalez and B. Holwerda and M. Franx and L. Spitler and R. Smit and D. Magee},
title = {UV LUMINOSITY FUNCTIONS AT REDSHIFTS z ∼ 4 TO z ∼ 10: 10,000 GALAXIES FROM HST LEGACY FIELDS*

†},
journal = {The Astrophysical Journal}
}

@ARTICLE{1987ApJ...319..730S,
       author = {{Solomon}, P.~M. and {Rivolo}, A.~R. and {Barrett}, J. and {Yahil}, A.},
        title = "{Mass, Luminosity, and Line Width Relations of Galactic Molecular Clouds}",
      journal = {\apj},
         year = 1987,
        month = aug,
       volume = {319},
        pages = {730},
          doi = {10.1086/165493},
       adsurl = {https://ui.adsabs.harvard.edu/abs/1987ApJ...319..730S}
}

@article{Tacconi_2013,
doi = {10.1088/0004-637X/768/1/74},
url = {https://dx.doi.org/10.1088/0004-637X/768/1/74},
year = {2013},
month = {apr},
publisher = {The American Astronomical Society},
volume = {768},
number = {1},
pages = {74},
author = {L. J. Tacconi and R. Neri and R. Genzel and F. Combes and A. Bolatto and M. C. Cooper and S. Wuyts and F. Bournaud and A. Burkert and J. Comerford and P. Cox and M. Davis and N. M. Förster Schreiber and S. García-Burillo and J. Gracia-Carpio and D. Lutz and T. Naab and S. Newman and A. Omont and A. Saintonge and K. Shapiro Griffin and A. Shapley and A. Sternberg and B. Weiner},
title = {PHIBSS: MOLECULAR GAS CONTENT AND SCALING RELATIONS IN z ∼ 1–3 MASSIVE, MAIN-SEQUENCE STAR-FORMING GALAXIES*},
journal = {The Astrophysical Journal}
}

@article{Aravena_2019,
doi = {10.3847/1538-4357/ab30df},
url = {https://dx.doi.org/10.3847/1538-4357/ab30df},
year = {2019},
month = {sep},
publisher = {The American Astronomical Society},
volume = {882},
number = {2},
pages = {136},
author = {Manuel Aravena and Roberto Decarli and Jorge Gónzalez-López and Leindert Boogaard and Fabian Walter and Chris Carilli and Gergö Popping and Axel Weiss and Roberto J. Assef and Roland Bacon and Franz Erik Bauer and Frank Bertoldi and Richard Bouwens and Thierry Contini and Paulo C. Cortes and Pierre Cox and Elisabete da Cunha and Emanuele Daddi and Tanio Díaz-Santos and David Elbaz and Jacqueline Hodge and Hanae Inami and Rob Ivison and Olivier Le Fèvre and Benjamin Magnelli and Pascal Oesch and Dominik Riechers and Ian Smail and Rachel S. Somerville and A. M. Swinbank and Bade Uzgil and Paul van der Werf and Jeff Wagg and Lutz Wisotzki},
title = {The ALMA Spectroscopic Survey in the Hubble Ultra Deep Field: Evolution of the Molecular Gas in CO-selected Galaxies},
journal = {The Astrophysical Journal}
}

@ARTICLE{2013MNRAS.429.3047B,
       author = {{Bothwell}, M.~S. and {Smail}, Ian and {Chapman}, S.~C. and {Genzel}, R. and {Ivison}, R.~J. and {Tacconi}, L.~J. and {Alaghband-Zadeh}, S. and {Bertoldi}, F. and {Blain}, A.~W. and {Casey}, C.~M. and {Cox}, P. and {Greve}, T.~R. and {Lutz}, D. and {Neri}, R. and {Omont}, A. and {Swinbank}, A.~M.},
        title = "{A survey of molecular gas in luminous sub-millimetre galaxies}",
      journal = {\mnras},
         year = 2013,
        month = mar,
       volume = {429},
       number = {4},
        pages = {3047-3067},
          doi = {10.1093/mnras/sts562},
archivePrefix = {arXiv},
       eprint = {1205.1511},
 primaryClass = {astro-ph.CO},
       adsurl = {https://ui.adsabs.harvard.edu/abs/2013MNRAS.429.3047B}
}

@ARTICLE{2018ApJ...853..179T,
       author = {{Tacconi}, L.~J. and {Genzel}, R. and {Saintonge}, A. and {Combes}, F. and {Garc{\'\i}a-Burillo}, S. and {Neri}, R. and {Bolatto}, A. and {Contini}, T. and {F{\"o}rster Schreiber}, N.~M. and {Lilly}, S. and {Lutz}, D. and {Wuyts}, S. and {Accurso}, G. and {Boissier}, J. and {Boone}, F. and {Bouch{\'e}}, N. and {Bournaud}, F. and {Burkert}, A. and {Carollo}, M. and {Cooper}, M. and {Cox}, P. and {Feruglio}, C. and {Freundlich}, J. and {Herrera-Camus}, R. and {Juneau}, S. and {Lippa}, M. and {Naab}, T. and {Renzini}, A. and {Salome}, P. and {Sternberg}, A. and {Tadaki}, K. and {{\"U}bler}, H. and {Walter}, F. and {Weiner}, B. and {Weiss}, A.},
        title = "{PHIBSS: Unified Scaling Relations of Gas Depletion Time and Molecular Gas Fractions}",
      journal = {\apj},
         year = 2018,
        month = feb,
       volume = {853},
       number = {2},
          eid = {179},
        pages = {179},
          doi = {10.3847/1538-4357/aaa4b4},
archivePrefix = {arXiv},
       eprint = {1702.01140},
 primaryClass = {astro-ph.GA},
       adsurl = {https://ui.adsabs.harvard.edu/abs/2018ApJ...853..179T}
}

@ARTICLE{2020A&A...643A...5D,
       author = {{Dessauges-Zavadsky}, M. and {Ginolfi}, M. and {Pozzi}, F. and {B{\'e}thermin}, M. and {Le F{\`e}vre}, O. and {Fujimoto}, S. and {Silverman}, J.~D. and {Jones}, G.~C. and {Vallini}, L. and {Schaerer}, D. and {Faisst}, A.~L. and {Khusanova}, Y. and {Fudamoto}, Y. and {Cassata}, P. and {Loiacono}, F. and {Capak}, P.~L. and {Yan}, L. and {Amorin}, R. and {Bardelli}, S. and {Boquien}, M. and {Cimatti}, A. and {Gruppioni}, C. and {Hathi}, N.~P. and {Ibar}, E. and {Koekemoer}, A.~M. and {Lemaux}, B.~C. and {Narayanan}, D. and {Oesch}, P.~A. and {Rodighiero}, G. and {Romano}, M. and {Talia}, M. and {Toft}, S. and {Vergani}, D. and {Zamorani}, G. and {Zucca}, E.},
        title = "{The ALPINE-ALMA [C II] survey. Molecular gas budget in the early Universe as traced by [C II]}",
      journal = {\aap},
         year = 2020,
        month = nov,
       volume = {643},
          eid = {A5},
        pages = {A5},
          doi = {10.1051/0004-6361/202038231},
archivePrefix = {arXiv},
       eprint = {2004.10771},
 primaryClass = {astro-ph.GA},
       adsurl = {https://ui.adsabs.harvard.edu/abs/2020A&A...643A...5D}
}

@ARTICLE{2024A&A...682A..24A,
       author = {{Aravena}, M. and {Heintz}, K. and {Dessauges-Zavadsky}, M. and {Oesch}, P. and {Algera}, H. and {Bouwens}, R. and {da Cunha}, E. and {Dayal}, P. and {De Looze}, I. and {Ferrara}, A. and {Fudamoto}, Y. and {Gonzalez}, V. and {Graziani}, L. and {Hygate}, A.~P.~S. and {Inami}, H. and {Pallottini}, A. and {Schneider}, R. and {Schouws}, S. and {Sommovigo}, L. and {Topping}, M. and {van der Werf}, P. and {Palla}, M.},
        title = "{The ALMA Reionization Era Bright Emission Line Survey: The molecular gas content of galaxies at z 7}",
      journal = {\aap},
         year = 2024,
        month = feb,
       volume = {682},
          eid = {A24},
        pages = {A24},
          doi = {10.1051/0004-6361/202347281},
archivePrefix = {arXiv},
       eprint = {2309.15948},
 primaryClass = {astro-ph.GA},
       adsurl = {https://ui.adsabs.harvard.edu/abs/2024A&A...682A..24A}
}

@ARTICLE{2020A&A...643A...2B,
       author = {{B{\'e}thermin}, M. and {Fudamoto}, Y. and {Ginolfi}, M. and {Loiacono}, F. and {Khusanova}, Y. and {Capak}, P.~L. and {Cassata}, P. and {Faisst}, A. and {Le F{\`e}vre}, O. and {Schaerer}, D. and {Silverman}, J.~D. and {Yan}, L. and {Amorin}, R. and {Bardelli}, S. and {Boquien}, M. and {Cimatti}, A. and {Davidzon}, I. and {Dessauges-Zavadsky}, M. and {Fujimoto}, S. and {Gruppioni}, C. and {Hathi}, N.~P. and {Ibar}, E. and {Jones}, G.~C. and {Koekemoer}, A.~M. and {Lagache}, G. and {Lemaux}, B.~C. and {Moreau}, C. and {Oesch}, P.~A. and {Pozzi}, F. and {Riechers}, D.~A. and {Talia}, M. and {Toft}, S. and {Vallini}, L. and {Vergani}, D. and {Zamorani}, G. and {Zucca}, E.},
        title = "{The ALPINE-ALMA [CII] survey: Data processing, catalogs, and statistical source properties}",
      journal = {\aap},
         year = 2020,
        month = nov,
       volume = {643},
          eid = {A2},
        pages = {A2},
          doi = {10.1051/0004-6361/202037649},
archivePrefix = {arXiv},
       eprint = {2002.00962},
 primaryClass = {astro-ph.GA},
       adsurl = {https://ui.adsabs.harvard.edu/abs/2020A&A...643A...2B}
}

@ARTICLE{2021MNRAS.505.4838L,
       author = {{Laporte}, N. and {Zitrin}, A. and {Ellis}, R.~S. and {Fujimoto}, S. and {Brammer}, G. and {Richard}, J. and {Oguri}, M. and {Caminha}, G.~B. and {Kohno}, K. and {Yoshimura}, Y. and {Ao}, Y. and {Bauer}, F.~E. and {Caputi}, K. and {Egami}, E. and {Espada}, D. and {Gonz{\'a}lez-L{\'o}pez}, J. and {Hatsukade}, B. and {Knudsen}, K.~K. and {Lee}, M.~M. and {Magdis}, G. and {Ouchi}, M. and {Valentino}, F. and {Wang}, T.},
        title = "{ALMA Lensing Cluster Survey: a strongly lensed multiply imaged dusty system at z {\ensuremath{\geq}} 6}",
      journal = {\mnras},
         year = 2021,
        month = aug,
       volume = {505},
       number = {4},
        pages = {4838-4846},
          doi = {10.1093/mnras/stab191},
archivePrefix = {arXiv},
       eprint = {2101.01740},
 primaryClass = {astro-ph.GA},
       adsurl = {https://ui.adsabs.harvard.edu/abs/2021MNRAS.505.4838L}
}

@ARTICLE{2017MNRAS.471.3305E,
       author = {{Ebeling}, H. and {Qi}, J. and {Richard}, J.},
        title = "{Fully stripped? The dynamics of dark and luminous matter in the massive cluster collision MACSJ0553.4-3342}",
      journal = {\mnras},
         year = 2017,
        month = nov,
       volume = {471},
       number = {3},
        pages = {3305-3322},
          doi = {10.1093/mnras/stx1636},
archivePrefix = {arXiv},
       eprint = {1706.03535},
 primaryClass = {astro-ph.CO},
       adsurl = {https://ui.adsabs.harvard.edu/abs/2017MNRAS.471.3305E}
}

@ARTICLE{2024ApJ...965..108U,
       author = {{Uematsu}, Ryosuke and {Ueda}, Yoshihiro and {Kohno}, Kotaro and {Toba}, Yoshiki and {Yamada}, Satoshi and {Smail}, Ian and {Umehata}, Hideki and {Fujimoto}, Seiji and {Hatsukade}, Bunyo and {Ao}, Yiping and {Bauer}, Franz Erik and {Brammer}, Gabriel and {Dessauges-Zavadsky}, Miroslava and {Espada}, Daniel and {Jolly}, Jean-Baptiste and {Koekemoer}, Anton M. and {Kokorev}, Vasily and {Magdis}, Georgios E. and {Oguri}, Masamune and {Sun}, Fengwu},
        title = "{ALMA Lensing Cluster Survey: Full Spectral Energy Distribution Analysis of z {\ensuremath{\sim}} 0.5{\textendash}6 Lensed Galaxies Detected with millimeter Observations}",
      journal = {\apj},
         year = 2024,
        month = apr,
       volume = {965},
       number = {2},
          eid = {108},
        pages = {108},
          doi = {10.3847/1538-4357/ad26f7},
       adsurl = {https://ui.adsabs.harvard.edu/abs/2024ApJ...965..108U}
}

@ARTICLE{2022ApJ...932...77S,
       author = {{Sun}, Fengwu and {Egami}, Eiichi and {Fujimoto}, Seiji and {Rawle}, Timothy and {Bauer}, Franz E. and {Kohno}, Kotaro and {Smail}, Ian and {P{\'e}rez-Gonz{\'a}lez}, Pablo G. and {Ao}, Yiping and {Chapman}, Scott C. and {Combes}, Francoise and {Dessauges-Zavadsky}, Miroslava and {Espada}, Daniel and {Gonz{\'a}lez-L{\'o}pez}, Jorge and {Koekemoer}, Anton M. and {Kokorev}, Vasily and {Lee}, Minju M. and {Morokuma-Matsui}, Kana and {Mu{\~n}oz Arancibia}, Alejandra M. and {Oguri}, Masamune and {Pell{\'o}}, Roser and {Ueda}, Yoshihiro and {Uematsu}, Ryosuke and {Valentino}, Francesco and {Van der Werf}, Paul and {Walth}, Gregory L. and {Zemcov}, Michael and {Zitrin}, Adi},
        title = "{ALMA Lensing Cluster Survey: ALMA-Herschel Joint Study of Lensed Dusty Star-forming Galaxies across z ≃ 0.5 - 6}",
      journal = {\apj},
         year = 2022,
        month = jun,
       volume = {932},
       number = {2},
          eid = {77},
        pages = {77},
          doi = {10.3847/1538-4357/ac6e3f},
archivePrefix = {arXiv},
       eprint = {2204.07187},
 primaryClass = {astro-ph.GA},
       adsurl = {https://ui.adsabs.harvard.edu/abs/2022ApJ...932...77S}
}

@ARTICLE{2018ApJ...864...49P,
       author = {{Pavesi}, Riccardo and {Sharon}, Chelsea E. and {Riechers}, Dominik A. and {Hodge}, Jacqueline A. and {Decarli}, Roberto and {Walter}, Fabian and {Carilli}, Chris L. and {Daddi}, Emanuele and {Smail}, Ian and {Dickinson}, Mark and {Ivison}, Rob J. and {Sargent}, Mark and {da Cunha}, Elisabete and {Aravena}, Manuel and {Darling}, Jeremy and {Smol{\v{c}}i{\'c}}, Vernesa and {Scoville}, Nicholas Z. and {Capak}, Peter L. and {Wagg}, Jeff},
        title = "{The CO Luminosity Density at High-z (COLDz) Survey: A Sensitive, Large-area Blind Search for Low-J CO Emission from Cold Gas in the Early Universe with the Karl G. Jansky Very Large Array}",
      journal = {\apj},
         year = 2018,
        month = sep,
       volume = {864},
       number = {1},
          eid = {49},
        pages = {49},
          doi = {10.3847/1538-4357/aacb79},
archivePrefix = {arXiv},
       eprint = {1808.04372},
 primaryClass = {astro-ph.GA},
       adsurl = {https://ui.adsabs.harvard.edu/abs/2018ApJ...864...49P}
}

@ARTICLE{2020ApJ...896L..21R,
       author = {{Riechers}, Dominik A. and {Boogaard}, Leindert A. and {Decarli}, Roberto and {Gonz{\'a}lez-L{\'o}pez}, Jorge and {Smail}, Ian and {Walter}, Fabian and {Aravena}, Manuel and {Carilli}, Christopher L. and {Cortes}, Paulo C. and {Cox}, Pierre and {D{\'\i}az-Santos}, Tanio and {Hodge}, Jacqueline A. and {Inami}, Hanae and {Ivison}, Rob J. and {Kaasinen}, Melanie and {Wagg}, Jeff and {Wei{\ss}}, Axel and {van der Werf}, Paul},
        title = "{VLA-ALMA Spectroscopic Survey in the Hubble Ultra Deep Field (VLASPECS): Total Cold Gas Masses and CO Line Ratios for z = 2-3 Main-sequence Galaxies}",
      journal = {\apjl},
         year = 2020,
        month = jun,
       volume = {896},
       number = {2},
          eid = {L21},
        pages = {L21},
          doi = {10.3847/2041-8213/ab9595},
archivePrefix = {arXiv},
       eprint = {2005.09653},
 primaryClass = {astro-ph.GA},
       adsurl = {https://ui.adsabs.harvard.edu/abs/2020ApJ...896L..21R}
}

@ARTICLE{2019ApJ...872....7R,
       author = {{Riechers}, Dominik A. and {Pavesi}, Riccardo and {Sharon}, Chelsea E. and {Hodge}, Jacqueline A. and {Decarli}, Roberto and {Walter}, Fabian and {Carilli}, Christopher L. and {Aravena}, Manuel and {da Cunha}, Elisabete and {Daddi}, Emanuele and {Dickinson}, Mark and {Smail}, Ian and {Capak}, Peter L. and {Ivison}, Rob J. and {Sargent}, Mark and {Scoville}, Nicholas Z. and {Wagg}, Jeff},
        title = "{COLDz: Shape of the CO Luminosity Function at High Redshift and the Cold Gas History of the Universe}",
      journal = {\apj},
         year = 2019,
        month = feb,
       volume = {872},
       number = {1},
          eid = {7},
        pages = {7},
          doi = {10.3847/1538-4357/aafc27},
archivePrefix = {arXiv},
       eprint = {1808.04371},
 primaryClass = {astro-ph.GA},
       adsurl = {https://ui.adsabs.harvard.edu/abs/2019ApJ...872....7R}
}

@ARTICLE{2020AJ....159..190L,
       author = {{Lenki{\'c}}, Laura and {Bolatto}, Alberto D. and {F{\"o}rster Schreiber}, Natascha M. and {Tacconi}, Linda J. and {Neri}, Roberto and {Combes}, Francoise and {Walter}, Fabian and {Garc{\'\i}a-Burillo}, Santiago and {Genzel}, Reinhard and {Lutz}, Dieter and {Cooper}, Michael C.},
        title = "{Plateau de Bure High-z Blue Sequence Survey 2 (PHIBSS2): Search for Secondary Sources, CO Luminosity Functions in the Field, and the Evolution of Molecular Gas Density through Cosmic Time}",
      journal = {\aj},
         year = 2020,
        month = may,
       volume = {159},
       number = {5},
          eid = {190},
        pages = {190},
          doi = {10.3847/1538-3881/ab7458},
archivePrefix = {arXiv},
       eprint = {1908.01791},
 primaryClass = {astro-ph.GA},
       adsurl = {https://ui.adsabs.harvard.edu/abs/2020AJ....159..190L}
}

@ARTICLE{2019ApJ...882..138D,
       author = {{Decarli}, Roberto and {Walter}, Fabian and {G{\'o}nzalez-L{\'o}pez}, Jorge and {Aravena}, Manuel and {Boogaard}, Leindert and {Carilli}, Chris and {Cox}, Pierre and {Daddi}, Emanuele and {Popping}, Gerg{\"o} and {Riechers}, Dominik and {Uzgil}, Bade and {Weiss}, Axel and {Assef}, Roberto J. and {Bacon}, Roland and {Bauer}, Franz Erik and {Bertoldi}, Frank and {Bouwens}, Rychard and {Contini}, Thierry and {Cortes}, Paulo C. and {da Cunha}, Elisabete and {D{\'\i}az-Santos}, Tanio and {Elbaz}, David and {Inami}, Hanae and {Hodge}, Jacqueline and {Ivison}, Rob and {Le F{\`e}vre}, Olivier and {Magnelli}, Benjamin and {Novak}, Mladen and {Oesch}, Pascal and {Rix}, Hans-Walter and {Sargent}, Mark T. and {Smail}, Ian and {Swinbank}, A. Mark and {Somerville}, Rachel S. and {van der Werf}, Paul and {Wagg}, Jeff and {Wisotzki}, Lutz},
        title = "{The ALMA Spectroscopic Survey in the HUDF: CO Luminosity Functions and the Molecular Gas Content of Galaxies through Cosmic History}",
      journal = {\apj},
         year = 2019,
        month = sep,
       volume = {882},
       number = {2},
          eid = {138},
        pages = {138},
          doi = {10.3847/1538-4357/ab30fe},
archivePrefix = {arXiv},
       eprint = {1903.09164},
 primaryClass = {astro-ph.GA},
       adsurl = {https://ui.adsabs.harvard.edu/abs/2019ApJ...882..138D}
}

@ARTICLE{2020ApJ...902..110D,
       author = {{Decarli}, Roberto and {Aravena}, Manuel and {Boogaard}, Leindert and {Carilli}, Chris and {Gonz{\'a}lez-L{\'o}pez}, Jorge and {Walter}, Fabian and {Cortes}, Paulo C. and {Cox}, Pierre and {da Cunha}, Elisabete and {Daddi}, Emanuele and {D{\'\i}az-Santos}, Tanio and {Hodge}, Jacqueline A. and {Inami}, Hanae and {Neeleman}, Marcel and {Novak}, Mladen and {Oesch}, Pascal and {Popping}, Gerg{\"o} and {Riechers}, Dominik and {Smail}, Ian and {Uzgil}, Bade and {van der Werf}, Paul and {Wagg}, Jeff and {Weiss}, Axel},
        title = "{The ALMA Spectroscopic Survey in the Hubble Ultra Deep Field: Multiband Constraints on Line-luminosity Functions and the Cosmic Density of Molecular Gas}",
      journal = {\apj},
         year = 2020,
        month = oct,
       volume = {902},
       number = {2},
          eid = {110},
        pages = {110},
          doi = {10.3847/1538-4357/abaa3b},
archivePrefix = {arXiv},
       eprint = {2009.10744},
 primaryClass = {astro-ph.GA},
       adsurl = {https://ui.adsabs.harvard.edu/abs/2020ApJ...902..110D}
}

@ARTICLE{2023ApJ...945..111B,
       author = {{Boogaard}, Leindert A. and {Decarli}, Roberto and {Walter}, Fabian and {Wei{\ss}}, Axel and {Popping}, Gerg{\"o} and {Neri}, Roberto and {Aravena}, Manuel and {Riechers}, Dominik and {Ellis}, Richard S. and {Carilli}, Chris and {Cox}, Pierre and {Pety}, J{\'e}r{\^o}me},
        title = "{A NOEMA Molecular Line Scan of the Hubble Deep Field North: Improved Constraints on the CO Luminosity Functions and Cosmic Density of Molecular Gas}",
      journal = {\apj},
         year = 2023,
        month = mar,
       volume = {945},
       number = {2},
          eid = {111},
        pages = {111},
          doi = {10.3847/1538-4357/acb4f0},
archivePrefix = {arXiv},
       eprint = {2301.05705},
 primaryClass = {astro-ph.GA},
       adsurl = {https://ui.adsabs.harvard.edu/abs/2023ApJ...945..111B}
}

@ARTICLE{2015ApJ...806..110D,
       author = {{da Cunha}, E. and {Walter}, F. and {Smail}, I.~R. and {Swinbank}, A.~M. and {Simpson}, J.~M. and {Decarli}, R. and {Hodge}, J.~A. and {Weiss}, A. and {van der Werf}, P.~P. and {Bertoldi}, F. and {Chapman}, S.~C. and {Cox}, P. and {Danielson}, A.~L.~R. and {Dannerbauer}, H. and {Greve}, T.~R. and {Ivison}, R.~J. and {Karim}, A. and {Thomson}, A.},
        title = "{An ALMA Survey of Sub-millimeter Galaxies in the Extended Chandra Deep Field South: Physical Properties Derived from Ultraviolet-to-radio Modeling}",
      journal = {\apj},
         year = 2015,
        month = jun,
       volume = {806},
       number = {1},
          eid = {110},
        pages = {110},
          doi = {10.1088/0004-637X/806/1/110},
archivePrefix = {arXiv},
       eprint = {1504.04376},
 primaryClass = {astro-ph.GA},
       adsurl = {https://ui.adsabs.harvard.edu/abs/2015ApJ...806..110D}
}

@ARTICLE{2022ApJ...935..167A,
       author = {{Astropy Collaboration} and {Price-Whelan}, Adrian M. and {Lim}, Pey Lian and {Earl}, Nicholas and {Starkman}, Nathaniel and {Bradley}, Larry and {Shupe}, David L. and {Patil}, Aarya A. and {Corrales}, Lia and {Brasseur}, C.~E. and {N{\"o}the}, Maximilian and {Donath}, Axel and {Tollerud}, Erik and {Morris}, Brett M. and {Ginsburg}, Adam and {Vaher}, Eero and {Weaver}, Benjamin A. and {Tocknell}, James and {Jamieson}, William and {van Kerkwijk}, Marten H. and {Robitaille}, Thomas P. and {Merry}, Bruce and {Bachetti}, Matteo and {G{\"u}nther}, H. Moritz and {Aldcroft}, Thomas L. and {Alvarado-Montes}, Jaime A. and {Archibald}, Anne M. and {B{\'o}di}, Attila and {Bapat}, Shreyas and {Barentsen}, Geert and {Baz{\'a}n}, Juanjo and {Biswas}, Manish and {Boquien}, M{\'e}d{\'e}ric and {Burke}, D.~J. and {Cara}, Daria and {Cara}, Mihai and {Conroy}, Kyle E. and {Conseil}, Simon and {Craig}, Matthew W. and {Cross}, Robert M. and {Cruz}, Kelle L. and {D'Eugenio}, Francesco and {Dencheva}, Nadia and {Devillepoix}, Hadrien A.~R. and {Dietrich}, J{\"o}rg P. and {Eigenbrot}, Arthur Davis and {Erben}, Thomas and {Ferreira}, Leonardo and {Foreman-Mackey}, Daniel and {Fox}, Ryan and {Freij}, Nabil and {Garg}, Suyog and {Geda}, Robel and {Glattly}, Lauren and {Gondhalekar}, Yash and {Gordon}, Karl D. and {Grant}, David and {Greenfield}, Perry and {Groener}, Austen M. and {Guest}, Steve and {Gurovich}, Sebastian and {Handberg}, Rasmus and {Hart}, Akeem and {Hatfield-Dodds}, Zac and {Homeier}, Derek and {Hosseinzadeh}, Griffin and {Jenness}, Tim and {Jones}, Craig K. and {Joseph}, Prajwel and {Kalmbach}, J. Bryce and {Karamehmetoglu}, Emir and {Ka{\l}uszy{\'n}ski}, Miko{\l}aj and {Kelley}, Michael S.~P. and {Kern}, Nicholas and {Kerzendorf}, Wolfgang E. and {Koch}, Eric W. and {Kulumani}, Shankar and {Lee}, Antony and {Ly}, Chun and {Ma}, Zhiyuan and {MacBride}, Conor and {Maljaars}, Jakob M. and {Muna}, Demitri and {Murphy}, N.~A. and {Norman}, Henrik and {O'Steen}, Richard and {Oman}, Kyle A. and {Pacifici}, Camilla and {Pascual}, Sergio and {Pascual-Granado}, J. and {Patil}, Rohit R. and {Perren}, Gabriel I. and {Pickering}, Timothy E. and {Rastogi}, Tanuj and {Roulston}, Benjamin R. and {Ryan}, Daniel F. and {Rykoff}, Eli S. and {Sabater}, Jose and {Sakurikar}, Parikshit and {Salgado}, Jes{\'u}s and {Sanghi}, Aniket and {Saunders}, Nicholas and {Savchenko}, Volodymyr and {Schwardt}, Ludwig and {Seifert-Eckert}, Michael and {Shih}, Albert Y. and {Jain}, Anany Shrey and {Shukla}, Gyanendra and {Sick}, Jonathan and {Simpson}, Chris and {Singanamalla}, Sudheesh and {Singer}, Leo P. and {Singhal}, Jaladh and {Sinha}, Manodeep and {Sip{\H{o}}cz}, Brigitta M. and {Spitler}, Lee R. and {Stansby}, David and {Streicher}, Ole and {{\v{S}}umak}, Jani and {Swinbank}, John D. and {Taranu}, Dan S. and {Tewary}, Nikita and {Tremblay}, Grant R. and {de Val-Borro}, Miguel and {Van Kooten}, Samuel J. and {Vasovi{\'c}}, Zlatan and {Verma}, Shresth and {de Miranda Cardoso}, Jos{\'e} Vin{\'\i}cius and {Williams}, Peter K.~G. and {Wilson}, Tom J. and {Winkel}, Benjamin and {Wood-Vasey}, W.~M. and {Xue}, Rui and {Yoachim}, Peter and {Zhang}, Chen and {Zonca}, Andrea and {Astropy Project Contributors}},
        title = "{The Astropy Project: Sustaining and Growing a Community-oriented Open-source Project and the Latest Major Release (v5.0) of the Core Package}",
      journal = {\apj},
         year = 2022,
        month = aug,
       volume = {935},
       number = {2},
          eid = {167},
        pages = {167},
          doi = {10.3847/1538-4357/ac7c74},
archivePrefix = {arXiv},
       eprint = {2206.14220},
 primaryClass = {astro-ph.IM},
       adsurl = {https://ui.adsabs.harvard.edu/abs/2022ApJ...935..167A}
}

@ARTICLE{2010PASJ...62.1017O,
       author = {{Oguri}, Masamune},
        title = "{The Mass Distribution of SDSS J1004+4112 Revisited}",
      journal = {\pasj},
         year = 2010,
        month = aug,
       volume = {62},
        pages = {1017},
          doi = {10.1093/pasj/62.4.1017},
archivePrefix = {arXiv},
       eprint = {1005.3103},
 primaryClass = {astro-ph.CO},
       adsurl = {https://ui.adsabs.harvard.edu/abs/2010PASJ...62.1017O}
}

@ARTICLE{2008ApJ...686.1503B,
       author = {{Brammer}, Gabriel B. and {van Dokkum}, Pieter G. and {Coppi}, Paolo},
        title = "{EAZY: A Fast, Public Photometric Redshift Code}",
      journal = {\apj},
         year = 2008,
        month = oct,
       volume = {686},
       number = {2},
        pages = {1503-1513},
          doi = {10.1086/591786},
archivePrefix = {arXiv},
       eprint = {0807.1533},
 primaryClass = {astro-ph},
       adsurl = {https://ui.adsabs.harvard.edu/abs/2008ApJ...686.1503B}
}

@ARTICLE{2024arXiv240609890T,
       author = {{Tsujita}, Akiyoshi and {Kohno}, Kotaro and {Huang}, Shuo and {Oguri}, Masamune and {Tadaki}, Ken-ichi and {Smail}, Ian and {Umehata}, Hideki and {Gao}, Zhen-Kai and {Wang}, Wei-Hao and {Sun}, Fengwu and {Fujimoto}, Seiji and {Wang}, Tao and {Uematsu}, Ryosuke and {Espada}, Daniel and {Valentino}, Francesco and {Ao}, Yiping and {Bauer}, Franz E. and {Hatsukade}, Bunyo and {Egusa}, Fumi and {Nishimura}, Yuri and {Koekemoer}, Anton M. and {Schaerer}, Daniel and {Lagos}, Claudia and {Dessauges-Zavadsky}, Miroslava and {Brammer}, Gabriel and {Caputi}, Karina and {Egami}, Eiichi and {Gonz{\'a}lez-L{\'o}pez}, Jorge and {Jolly}, Jean-Baptiste and {Knudsen}, Kirsten K. and {Kokorev}, Vasily and {Magdis}, Georgios E. and {Ouchi}, Masami and {Toft}, Sune and {Wu}, John F. and {Zitrin}, Adi},
        title = "{ALMA Lensing Cluster Survey: Physical characterization of near-infrared-dark intrinsically faint ALMA sources at z=2-4}",
      journal = {arXiv e-prints},
         year = 2024,
        month = jun,
          eid = {arXiv:2406.09890},
        pages = {arXiv:2406.09890},
          doi = {10.48550/arXiv.2406.09890},
archivePrefix = {arXiv},
       eprint = {2406.09890},
 primaryClass = {astro-ph.GA},
       adsurl = {https://ui.adsabs.harvard.edu/abs/2024arXiv240609890T}
}

@ARTICLE{2017A&A...608A.138G,
       author = {{Gonz{\'a}lez-L{\'o}pez}, J. and {Bauer}, F.~E. and {Aravena}, M. and {Laporte}, N. and {Bradley}, L. and {Carrasco}, M. and {Carvajal}, R. and {Demarco}, R. and {Infante}, L. and {Kneissl}, R. and {Koekemoer}, A.~M. and {Mu{\~n}oz Arancibia}, A.~M. and {Troncoso}, P. and {Villard}, E. and {Zitrin}, A.},
        title = "{The ALMA Frontier Fields Survey. III. 1.1 mm emission line identifications in Abell 2744, MACSJ 0416.1-2403, MACSJ 1149.5+2223, Abell 370, and Abell S1063}",
      journal = {\aap},
         year = 2017,
        month = dec,
       volume = {608},
          eid = {A138},
        pages = {A138},
          doi = {10.1051/0004-6361/201730961},
archivePrefix = {arXiv},
       eprint = {1704.03007},
 primaryClass = {astro-ph.GA},
       adsurl = {https://ui.adsabs.harvard.edu/abs/2017A&A...608A.138G}
}

@ARTICLE{2015A&A...575A..74S,
       author = {{Schreiber}, C. and {Pannella}, M. and {Elbaz}, D. and {B{\'e}thermin}, M. and {Inami}, H. and {Dickinson}, M. and {Magnelli}, B. and {Wang}, T. and {Aussel}, H. and {Daddi}, E. and {Juneau}, S. and {Shu}, X. and {Sargent}, M.~T. and {Buat}, V. and {Faber}, S.~M. and {Ferguson}, H.~C. and {Giavalisco}, M. and {Koekemoer}, A.~M. and {Magdis}, G. and {Morrison}, G.~E. and {Papovich}, C. and {Santini}, P. and {Scott}, D.},
        title = "{The Herschel view of the dominant mode of galaxy growth from z = 4 to the present day}",
      journal = {\aap},
         year = 2015,
        month = mar,
       volume = {575},
          eid = {A74},
        pages = {A74},
          doi = {10.1051/0004-6361/201425017},
archivePrefix = {arXiv},
       eprint = {1409.5433},
 primaryClass = {astro-ph.GA},
       adsurl = {https://ui.adsabs.harvard.edu/abs/2015A&A...575A..74S}
}

@ARTICLE{2013ARA&A..51..105C,
       author = {{Carilli}, C.~L. and {Walter}, F.},
        title = "{Cool Gas in High-Redshift Galaxies}",
      journal = {\araa},
         year = 2013,
        month = aug,
       volume = {51},
       number = {1},
        pages = {105-161},
          doi = {10.1146/annurev-astro-082812-140953},
archivePrefix = {arXiv},
       eprint = {1301.0371},
 primaryClass = {astro-ph.CO},
       adsurl = {https://ui.adsabs.harvard.edu/abs/2013ARA&A..51..105C}
}

@ARTICLE{2014ApJ...782...78D,
       author = {{Decarli}, R. and {Walter}, F. and {Carilli}, C. and {Riechers}, D. and {Cox}, P. and {Neri}, R. and {Aravena}, M. and {Bell}, E. and {Bertoldi}, F. and {Colombo}, D. and {Da Cunha}, E. and {Daddi}, E. and {Dickinson}, M. and {Downes}, D. and {Ellis}, R. and {Lentati}, L. and {Maiolino}, R. and {Menten}, K.~M. and {Rix}, H. -W. and {Sargent}, M. and {Stark}, D. and {Weiner}, B. and {Weiss}, A.},
        title = "{A Molecular Line Scan in the Hubble Deep Field North}",
      journal = {\apj},
         year = 2014,
        month = feb,
       volume = {782},
       number = {2},
          eid = {78},
        pages = {78},
          doi = {10.1088/0004-637X/782/2/78},
archivePrefix = {arXiv},
       eprint = {1312.6364},
 primaryClass = {astro-ph.GA},
       adsurl = {https://ui.adsabs.harvard.edu/abs/2014ApJ...782...78D}
}

@ARTICLE{2014ApJ...782...79W,
       author = {{Walter}, F. and {Decarli}, R. and {Sargent}, M. and {Carilli}, C. and {Dickinson}, M. and {Riechers}, D. and {Ellis}, R. and {Stark}, D. and {Weiner}, B. and {Aravena}, M. and {Bell}, E. and {Bertoldi}, F. and {Cox}, P. and {Da Cunha}, E. and {Daddi}, E. and {Downes}, D. and {Lentati}, L. and {Maiolino}, R. and {Menten}, K.~M. and {Neri}, R. and {Rix}, H. -W. and {Weiss}, A.},
        title = "{A Molecular Line Scan in the Hubble Deep Field North: Constraints on the CO Luminosity Function and the Cosmic H$_{2}$ Density}",
      journal = {\apj},
         year = 2014,
        month = feb,
       volume = {782},
       number = {2},
          eid = {79},
        pages = {79},
          doi = {10.1088/0004-637X/782/2/79},
archivePrefix = {arXiv},
       eprint = {1312.6365},
 primaryClass = {astro-ph.CO},
       adsurl = {https://ui.adsabs.harvard.edu/abs/2014ApJ...782...79W}
}

@ARTICLE{2014ApJ...788..147R,
       author = {{Rangwala}, Naseem and {Maloney}, Philip R. and {Glenn}, Jason and {Wilson}, Christine D. and {Kamenetzky}, Julia and {Schirm}, Maximilien R.~P. and {Spinoglio}, Luigi and {Pereira Santaella}, Miguel},
        title = "{First Extragalactic Detection of Submillimeter CH Rotational Lines from the Herschel Space Observatory}",
      journal = {\apj},
         year = 2014,
        month = jun,
       volume = {788},
       number = {2},
          eid = {147},
        pages = {147},
          doi = {10.1088/0004-637X/788/2/147},
archivePrefix = {arXiv},
       eprint = {1404.7200},
 primaryClass = {astro-ph.GA},
       adsurl = {https://ui.adsabs.harvard.edu/abs/2014ApJ...788..147R}
}

@ARTICLE{2014ApJ...785..149S,
       author = {{Spilker}, J.~S. and {Marrone}, D.~P. and {Aguirre}, J.~E. and {Aravena}, M. and {Ashby}, M.~L.~N. and {B{\'e}thermin}, M. and {Bradford}, C.~M. and {Bothwell}, M.~S. and {Brodwin}, M. and {Carlstrom}, J.~E. and {Chapman}, S.~C. and {Crawford}, T.~M. and {de Breuck}, C. and {Fassnacht}, C.~D. and {Gonzalez}, A.~H. and {Greve}, T.~R. and {Gullberg}, B. and {Hezaveh}, Y. and {Holzapfel}, W.~L. and {Husband}, K. and {Ma}, J. and {Malkan}, M. and {Murphy}, E.~J. and {Reichardt}, C.~L. and {Rotermund}, K.~M. and {Stalder}, B. and {Stark}, A.~A. and {Strandet}, M. and {Vieira}, J.~D. and {Wei{\ss}}, A. and {Welikala}, N.},
        title = "{The Rest-frame Submillimeter Spectrum of High-redshift, Dusty, Star-forming Galaxies}",
      journal = {\apj},
         year = 2014,
        month = apr,
       volume = {785},
       number = {2},
          eid = {149},
        pages = {149},
          doi = {10.1088/0004-637X/785/2/149},
archivePrefix = {arXiv},
       eprint = {1403.1667},
 primaryClass = {astro-ph.GA},
       adsurl = {https://ui.adsabs.harvard.edu/abs/2014ApJ...785..149S}
}

@ARTICLE{2024arXiv240218543F,
       author = {{Fujimoto}, S. and {Ouchi}, M. and {Kohno}, K. and {Valentino}, F. and {Gim\textbackslash'enez-Arteaga}, C. and {Brammer}, G.~B. and {Furtak}, L.~J. and {Kohandel}, M. and {Oguri}, M. and {Pallottini}, A. and {Richard}, J. and {Zitrin}, A. and {Bauer}, F.~E. and {Boylan-Kolchin}, M. and {Dessauges-Zavadsky}, M. and {Egami}, E. and {Finkelstein}, S.~L. and {Ma}, Z. and {Smail}, I. and {Watson}, D. and {Hutchison}, T.~A. and {Rigby}, J.~R. and {Welch}, B.~D. and {Ao}, Y. and {Bradley}, L.~D. and {Caminha}, G.~B. and {Caputi}, K.~I. and {Espada}, D. and {Endsley}, R. and {Fudamoto}, Y. and {Gonz\textbackslash'alez-L\textbackslash'opez}, J. and {Hatsukade}, B. and {Koekemoer}, A.~M. and {Kokorev}, V. and {Laporte}, N. and {Lee}, M. and {Magdis}, G.~E. and {Ono}, Y. and {Rizzo}, F. and {Shibuya}, T. and {Shimasaku}, K. and {Sun}, F. and {Toft}, S. and {Umehata}, H. and {Wang}, T. and {Yajima}, H.},
        title = "{Primordial Rotating Disk Composed of $\geq$15 Dense Star-Forming Clumps at Cosmic Dawn}",
      journal = {arXiv e-prints},
         year = {2024b},
        month = feb,
          eid = {arXiv:2402.18543},
        pages = {arXiv:2402.18543},
          doi = {10.48550/arXiv.2402.18543},
archivePrefix = {arXiv},
       eprint = {2402.18543},
 primaryClass = {astro-ph.GA},
       adsurl = {https://ui.adsabs.harvard.edu/abs/2024arXiv240218543F}
}

@ARTICLE{2013ARA&A..51..207B,
       author = {{Bolatto}, Alberto D. and {Wolfire}, Mark and {Leroy}, Adam K.},
        title = "{The CO-to-H$_{2}$ Conversion Factor}",
      journal = {\araa},
         year = 2013,
        month = aug,
       volume = {51},
       number = {1},
        pages = {207-268},
          doi = {10.1146/annurev-astro-082812-140944},
archivePrefix = {arXiv},
       eprint = {1301.3498},
 primaryClass = {astro-ph.GA},
       adsurl = {https://ui.adsabs.harvard.edu/abs/2013ARA&A..51..207B}
}

@ARTICLE{2008ApJ...679..481P,
       author = {{Pineda}, Jaime E. and {Caselli}, Paola and {Goodman}, Alyssa A.},
        title = "{CO Isotopologues in the Perseus Molecular Cloud Complex: the X-factor and Regional Variations}",
      journal = {\apj},
         year = 2008,
        month = may,
       volume = {679},
       number = {1},
        pages = {481-496},
          doi = {10.1086/586883},
archivePrefix = {arXiv},
       eprint = {0802.0708},
 primaryClass = {astro-ph},
       adsurl = {https://ui.adsabs.harvard.edu/abs/2008ApJ...679..481P}
}

@ARTICLE{2023ApJ...948...44R,
       author = {{Reuter}, C. and {Spilker}, J.~S. and {Vieira}, J.~D. and {Marrone}, D.~P. and {Weiss}, A. and {Aravena}, M. and {Archipley}, M.~A. and {Chapman}, S.~C. and {Gonzalez}, A. and {Greve}, T.~R. and {Hayward}, C.~C. and {Hill}, R. and {Jarugula}, S. and {Kim}, S. and {Malkan}, M. and {Phadke}, K.~A. and {Stark}, A.~A. and {Sulzenauer}, N. and {Vizgan}, D.},
        title = "{The Rest-frame Submillimeter Spectrum of High-redshift, Dusty, Star-forming Galaxies from the SPT-SZ Survey}",
      journal = {\apj},
         year = 2023,
        month = may,
       volume = {948},
       number = {1},
          eid = {44},
        pages = {44},
          doi = {10.3847/1538-4357/acaf51},
archivePrefix = {arXiv},
       eprint = {2210.11671},
 primaryClass = {astro-ph.GA},
       adsurl = {https://ui.adsabs.harvard.edu/abs/2023ApJ...948...44R}
}

@ARTICLE{2001A&A...370L..49M,
       author = {{M{\"u}ller}, H.~S.~P. and {Thorwirth}, S. and {Roth}, D.~A. and {Winnewisser}, G.},
        title = "{The Cologne Database for Molecular Spectroscopy, CDMS}",
      journal = {\aap},
         year = 2001,
        month = apr,
       volume = {370},
        pages = {L49-L52},
          doi = {10.1051/0004-6361:20010367},
       adsurl = {https://ui.adsabs.harvard.edu/abs/2001A&A...370L..49M}
}

@ARTICLE{2005JMoSt.742..215M,
       author = {{M{\"u}ller}, Holger S.~P. and {Schl{\"o}der}, Frank and {Stutzki}, J{\"u}rgen and {Winnewisser}, Gisbert},
        title = "{The Cologne Database for Molecular Spectroscopy, CDMS: a useful tool for astronomers and spectroscopists}",
      journal = {Journal of Molecular Structure},
         year = 2005,
        month = may,
       volume = {742},
       number = {1-3},
        pages = {215-227},
          doi = {10.1016/j.molstruc.2005.01.027},
       adsurl = {https://ui.adsabs.harvard.edu/abs/2005JMoSt.742..215M}
}

@ARTICLE{1998JQSRT..60..883P,
       author = {{Pickett}, H.~M. and {Poynter}, R.~L. and {Cohen}, E.~A. and {Delitsky}, M.~L. and {Pearson}, J.~C. and {M{\"u}ller}, H.~S.~P.},
        title = "{Submillimeter, millimeter and microwave spectral line catalog.}",
      journal = {\jqsrt},
         year = 1998,
        month = nov,
       volume = {60},
       number = {5},
        pages = {883-890},
          doi = {10.1016/S0022-4073(98)00091-0},
       adsurl = {https://ui.adsabs.harvard.edu/abs/1998JQSRT..60..883P}
}

@ARTICLE{2024A&A...686A..63G,
       author = {{Gim{\'e}nez-Arteaga}, C. and {Fujimoto}, S. and {Valentino}, F. and {Brammer}, G.~B. and {Mason}, C.~A. and {Rizzo}, F. and {Rusakov}, V. and {Colina}, L. and {Prieto-Lyon}, G. and {Oesch}, P.~A. and {Espada}, D. and {Heintz}, K.~E. and {Knudsen}, K.~K. and {Dessauges-Zavadsky}, M. and {Laporte}, N. and {Lee}, M. and {Magdis}, G.~E. and {Ono}, Y. and {Ao}, Y. and {Ouchi}, M. and {Kohno}, K. and {Koekemoer}, A.~M.},
        title = "{Outshining in the spatially resolved analysis of a strongly lensed galaxy at z = 6.072 with JWST NIRCam}",
      journal = {\aap},
         year = 2024,
        month = jun,
       volume = {686},
          eid = {A63},
        pages = {A63},
          doi = {10.1051/0004-6361/202349135},
archivePrefix = {arXiv},
       eprint = {2402.17875},
 primaryClass = {astro-ph.GA},
       adsurl = {https://ui.adsabs.harvard.edu/abs/2024A&A...686A..63G}
}

@ARTICLE{2024A&A...685A.138V,
       author = {{Valentino}, F. and {Fujimoto}, S. and {Gim{\'e}nez-Arteaga}, C. and {Brammer}, G. and {Kohno}, K. and {Sun}, F. and {Kokorev}, V. and {Bauer}, F.~E. and {Di Cesare}, C. and {Espada}, D. and {Lee}, M. and {Dessauges-Zavadsky}, M. and {Ao}, Y. and {Koekemoer}, A.~M. and {Ouchi}, M. and {Wu}, J.~F. and {Egami}, E. and {Jolly}, J. -B. and {Lagos}, C. del P. and {Magdis}, G.~E. and {Schaerer}, D. and {Shimasaku}, K. and {Umehata}, H. and {Wang}, W. -H.},
        title = "{The cold interstellar medium of a normal sub-L$^{{\ensuremath{\star}}}$ galaxy at the end of reionization}",
      journal = {\aap},
         year = 2024,
        month = may,
       volume = {685},
          eid = {A138},
        pages = {A138},
          doi = {10.1051/0004-6361/202348128},
archivePrefix = {arXiv},
       eprint = {2402.17845},
 primaryClass = {astro-ph.GA},
       adsurl = {https://ui.adsabs.harvard.edu/abs/2024A&A...685A.138V}
}

@ARTICLE{2021ApJ...909...56L,
       author = {{Liu}, Daizhong and {Daddi}, Emanuele and {Schinnerer}, Eva and {Saito}, Toshiki and {Leroy}, Adam and {Silverman}, John D. and {Valentino}, Francesco and {Magdis}, Georgios E. and {Gao}, Yu and {Jin}, Shuowen and {Puglisi}, Annagrazia and {Groves}, Brent},
        title = "{CO Excitation, Molecular Gas Density, and Interstellar Radiation Field in Local and High-redshift Galaxies}",
      journal = {\apj},
         year = 2021,
        month = mar,
       volume = {909},
       number = {1},
          eid = {56},
        pages = {56},
          doi = {10.3847/1538-4357/abd801},
archivePrefix = {arXiv},
       eprint = {2101.06646},
 primaryClass = {astro-ph.GA},
       adsurl = {https://ui.adsabs.harvard.edu/abs/2021ApJ...909...56L}
}

@ARTICLE{2020ApJ...902..109B,
       author = {{Boogaard}, Leindert A. and {van der Werf}, Paul and {Weiss}, Axel and {Popping}, Gerg{\"o} and {Decarli}, Roberto and {Walter}, Fabian and {Aravena}, Manuel and {Bouwens}, Rychard and {Riechers}, Dominik and {Gonz{\'a}lez-L{\'o}pez}, Jorge and {Smail}, Ian and {Carilli}, Chris and {Kaasinen}, Melanie and {Daddi}, Emanuele and {Cox}, Pierre and {D{\'\i}az-Santos}, Tanio and {Inami}, Hanae and {Cortes}, Paulo C. and {Wagg}, Jeff},
        title = "{The ALMA Spectroscopic Survey in the Hubble Ultra Deep Field: CO Excitation and Atomic Carbon in Star-forming Galaxies at z = 1-3}",
      journal = {\apj},
         year = 2020,
        month = oct,
       volume = {902},
       number = {2},
          eid = {109},
        pages = {109},
          doi = {10.3847/1538-4357/abb82f},
archivePrefix = {arXiv},
       eprint = {2009.04348},
 primaryClass = {astro-ph.GA},
       adsurl = {https://ui.adsabs.harvard.edu/abs/2020ApJ...902..109B}
}

@ARTICLE{2025A&A...693A.119C,
       author = {{Casavecchia}, Benedetta and {Maio}, Umberto and {P{\'e}roux}, C{\'e}line and {Ciardi}, Benedetta},
        title = "{Atomic and molecular gas as traced by [C II] emission}",
      journal = {\aap},
         year = 2025,
        month = jan,
       volume = {693},
          eid = {A119},
        pages = {A119},
          doi = {10.1051/0004-6361/202452282},
archivePrefix = {arXiv},
       eprint = {2410.14284},
 primaryClass = {astro-ph.GA},
       adsurl = {https://ui.adsabs.harvard.edu/abs/2025A&A...693A.119C}
}

@ARTICLE{2012ApJ...758L...9M,
       author = {{Magdis}, Georgios E. and {Daddi}, E. and {Sargent}, M. and {Elbaz}, D. and {Gobat}, R. and {Dannerbauer}, H. and {Feruglio}, C. and {Tan}, Q. and {Rigopoulou}, D. and {Charmandaris}, V. and {Dickinson}, M. and {Reddy}, N. and {Aussel}, H.},
        title = "{The Molecular Gas Content of z = 3 Lyman Break Galaxies: Evidence of a Non-evolving Gas Fraction in Main-sequence Galaxies at z > 2}",
      journal = {\apjl},
         year = 2012,
        month = oct,
       volume = {758},
       number = {1},
          eid = {L9},
        pages = {L9},
          doi = {10.1088/2041-8205/758/1/L9},
archivePrefix = {arXiv},
       eprint = {1209.1484},
 primaryClass = {astro-ph.CO},
       adsurl = {https://ui.adsabs.harvard.edu/abs/2012ApJ...758L...9M}
}

@ARTICLE{2016ApJ...820...83S,
       author = {{Scoville}, N. and {Sheth}, K. and {Aussel}, H. and {Vanden Bout}, P. and {Capak}, P. and {Bongiorno}, A. and {Casey}, C.~M. and {Murchikova}, L. and {Koda}, J. and {{\'A}lvarez-M{\'a}rquez}, J. and {Lee}, N. and {Laigle}, C. and {McCracken}, H.~J. and {Ilbert}, O. and {Pope}, A. and {Sanders}, D. and {Chu}, J. and {Toft}, S. and {Ivison}, R.~J. and {Manohar}, S.},
        title = "{ISM Masses and the Star formation Law at Z = 1 to 6: ALMA Observations of Dust Continuum in 145 Galaxies in the COSMOS Survey Field}",
      journal = {\apj},
         year = 2016,
        month = apr,
       volume = {820},
       number = {2},
          eid = {83},
        pages = {83},
          doi = {10.3847/0004-637X/820/2/83},
archivePrefix = {arXiv},
       eprint = {1511.05149},
 primaryClass = {astro-ph.GA},
       adsurl = {https://ui.adsabs.harvard.edu/abs/2016ApJ...820...83S}
}

@ARTICLE{2005MNRAS.363....2K,
       author = {{Kere{\v{s}}}, Du{\v{s}}an and {Katz}, Neal and {Weinberg}, David H. and {Dav{\'e}}, Romeel},
        title = "{How do galaxies get their gas?}",
      journal = {\mnras},
         year = 2005,
        month = oct,
       volume = {363},
       number = {1},
        pages = {2-28},
          doi = {10.1111/j.1365-2966.2005.09451.x},
archivePrefix = {arXiv},
       eprint = {astro-ph/0407095},
 primaryClass = {astro-ph},
       adsurl = {https://ui.adsabs.harvard.edu/abs/2005MNRAS.363....2K}
}

@ARTICLE{2009Natur.457..451D,
       author = {{Dekel}, A. and {Birnboim}, Y. and {Engel}, G. and {Freundlich}, J. and {Goerdt}, T. and {Mumcuoglu}, M. and {Neistein}, E. and {Pichon}, C. and {Teyssier}, R. and {Zinger}, E.},
        title = "{Cold streams in early massive hot haloes as the main mode of galaxy formation}",
      journal = {\nat},
         year = 2009,
        month = jan,
       volume = {457},
       number = {7228},
        pages = {451-454},
          doi = {10.1038/nature07648},
archivePrefix = {arXiv},
       eprint = {0808.0553},
 primaryClass = {astro-ph},
       adsurl = {https://ui.adsabs.harvard.edu/abs/2009Natur.457..451D}
}

@ARTICLE{2023ApJ...952...81F,
       author = {{Frye}, Brenda L. and {Pascale}, Massimo and {Foo}, Nicholas and {Leimbach}, Reagen and {Garuda}, Nikhil and {Robles}, Paulina Soto and {Summers}, Jake and {Diaz}, Carlos and {Kamieneski}, Patrick and {Furtak}, Lukas J. and {Cohen}, Seth H. and {Diego}, Jose and {Beauchesne}, Benjamin and {Windhorst}, Rogier A. and {Willner}, S.~P. and {Koekemoer}, Anton M. and {Zitrin}, Adi and {Caminha}, Gabriel and {Caputi}, Karina I. and {Coe}, Dan and {Conselice}, Christopher J. and {Dai}, Liang and {Dole}, Herv{\'e} and {Driver}, Simon P. and {Grogin}, Norman A. and {Harrington}, Kevin and {Jansen}, Rolf A. and {Kneib}, Jean-Paul and {Lehnert}, Matt and {Lowenthal}, James and {Marshall}, Madeline A. and {Menanteau}, Felipe and {Pampliega}, Bel{\'e}n Alcalde and {Pirzkal}, Nor and {Polletta}, Maria del Carmen and {Richard}, Johan and {Robotham}, Aaron and {Ryan}, Russell E. and {Rutkowski}, Michael J. and {Sif{\'o}n}, Christ{\'o}bal and {Tompkins}, Scott and {Wang}, Daniel and {Yan}, Haojing and {Yun}, Min S.},
        title = "{The JWST PEARLS View of the El Gordo Galaxy Cluster and of the Structure It Magnifies}",
      journal = {\apj},
         year = 2023,
        month = jul,
       volume = {952},
       number = {1},
          eid = {81},
        pages = {81},
          doi = {10.3847/1538-4357/acd929},
archivePrefix = {arXiv},
       eprint = {2303.03556},
 primaryClass = {astro-ph.GA},
       adsurl = {https://ui.adsabs.harvard.edu/abs/2023ApJ...952...81F}
}

@ARTICLE{2025ApJ...983L..22K,
       author = {{Kokorev}, Vasily and {Atek}, Hakim and {Chisholm}, John and {Endsley}, Ryan and {Chemerynska}, Iryna and {Mu{\~n}oz}, Julian B. and {Furtak}, Lukas J. and {Pan}, Richard and {Berg}, Danielle and {Fujimoto}, Seiji and {Oesch}, Pascal A. and {Weibel}, Andrea and {Adamo}, Angela and {Blaizot}, Jeremy and {Bouwens}, Rychard and {Dessauges-Zavadsky}, Miroslava and {Khullar}, Gourav and {Korber}, Damien and {Goovaerts}, Ilias and {Jecmen}, Michelle and {Labb{\'e}}, Ivo and {Leclercq}, Floriane and {Marques-Chaves}, Rui and {Mason}, Charlotte and {McQuinn}, Kristen B.~W. and {Naidu}, Rohan and {Natarajan}, Priyamvada and {Nelson}, Erica and {Rosdahl}, Joki and {Saldana-Lopez}, Alberto and {Schaerer}, Daniel and {Trebitsch}, Maxime and {Volonteri}, Marta and {Zitrin}, Adi},
        title = "{A Glimpse of the New Redshift Frontier through AS1063}",
      journal = {\apjl},
         year = 2025,
        month = apr,
       volume = {983},
       number = {1},
          eid = {L22},
        pages = {L22},
          doi = {10.3847/2041-8213/adc458},
archivePrefix = {arXiv},
       eprint = {2411.13640},
 primaryClass = {astro-ph.GA},
       adsurl = {https://ui.adsabs.harvard.edu/abs/2025ApJ...983L..22K}
}

@ARTICLE{2020MNRAS.496.2591O,
       author = {{Okabe}, Taizo and {Oguri}, Masamune and {Peirani}, S{\'e}bastien and {Suto}, Yasushi and {Dubois}, Yohan and {Pichon}, Christophe and {Kitayama}, Tetsu and {Sasaki}, Shin and {Nishimichi}, Takahiro},
        title = "{Shapes and alignments of dark matter haloes and their brightest cluster galaxies in 39 strong lensing clusters}",
      journal = {\mnras},
         year = 2020,
        month = aug,
       volume = {496},
       number = {3},
        pages = {2591-2604},
          doi = {10.1093/mnras/staa1479},
archivePrefix = {arXiv},
       eprint = {2005.11469},
 primaryClass = {astro-ph.CO},
       adsurl = {https://ui.adsabs.harvard.edu/abs/2020MNRAS.496.2591O}
}

@ARTICLE{1973ApL....15...79B,
       author = {{Black}, J.~H. and {Dalgarno}, A.},
        title = "{The Formation of CH in Interstellar Clouds}",
      journal = {\aplett},
         year = 1973,
        month = oct,
       volume = {15},
        pages = {79},
       adsurl = {https://ui.adsabs.harvard.edu/abs/1973ApL....15...79B}
}

@ARTICLE{2012ApJ...758..108S,
       author = {{Spinoglio}, Luigi and {Pereira-Santaella}, Miguel and {Busquet}, Gemma and {Schirm}, Maximilien R.~P. and {Wilson}, Christine D. and {Glenn}, Jason and {Kamenetzky}, Julia and {Rangwala}, Naseem and {Maloney}, Philip R. and {Parkin}, Tara J. and {Bendo}, George J. and {Madden}, Suzanne C. and {Wolfire}, Mark G. and {Boselli}, Alessandro and {Cooray}, Asantha and {Page}, Mathew J.},
        title = "{Submillimeter Line Spectrum of the Seyfert Galaxy NGC 1068 from the Herschel-SPIRE Fourier Transform Spectrometer}",
      journal = {\apj},
         year = 2012,
        month = oct,
       volume = {758},
       number = {2},
          eid = {108},
        pages = {108},
          doi = {10.1088/0004-637X/758/2/108},
archivePrefix = {arXiv},
       eprint = {1208.6132},
 primaryClass = {astro-ph.GA},
       adsurl = {https://ui.adsabs.harvard.edu/abs/2012ApJ...758..108S}
}

@ARTICLE{2015PASP..127..266M,
       author = {{Mangum}, Jeffrey G. and {Shirley}, Yancy L.},
        title = "{How to Calculate Molecular Column Density}",
      journal = {\pasp},
         year = 2015,
        month = mar,
       volume = {127},
       number = {949},
        pages = {266},
          doi = {10.1086/680323},
archivePrefix = {arXiv},
       eprint = {1501.01703},
 primaryClass = {astro-ph.IM},
       adsurl = {https://ui.adsabs.harvard.edu/abs/2015PASP..127..266M}
}

@ARTICLE{2024ApJS..275...36F,
       author = {{Fujimoto}, Seiji and {Kohno}, Kotaro and {Ouchi}, Masami and {Oguri}, Masamune and {Kokorev}, Vasily and {Brammer}, Gabriel and {Sun}, Fengwu and {Gonz{\'a}lez-L{\'o}pez}, Jorge and {Bauer}, Franz E. and {Caminha}, Gabriel B. and {Hatsukade}, Bunyo and {Richard}, Johan and {Smail}, Ian and {Tsujita}, Akiyoshi and {Ueda}, Yoshihiro and {Uematsu}, Ryosuke and {Zitrin}, Adi and {Coe}, Dan and {Kneib}, Jean-Paul and {Postman}, Marc and {Umetsu}, Keiichi and {Lagos}, Claudia del P. and {Popping}, Gerg{\"o} and {Ao}, Yiping and {Bradley}, Larry and {Caputi}, Karina and {Dessauges-Zavadsky}, Miroslava and {Egami}, Eiichi and {Espada}, Daniel and {Ivison}, R.~J. and {Jauzac}, Mathilde and {Knudsen}, Kirsten K. and {Koekemoer}, Anton M. and {Magdis}, Georgios E. and {Mahler}, Guillaume and {Mu{\~n}oz Arancibia}, A.~M. and {Rawle}, Timothy and {Shimasaku}, Kazuhiro and {Toft}, Sune and {Umehata}, Hideki and {Valentino}, Francesco and {Wang}, Tao and {Wang}, Wei-Hao},
        title = "{ALMA Lensing Cluster Survey: Deep 1.2 mm Number Counts and Infrared Luminosity Functions at z ≃ 1{\textendash}8}",
      journal = {\apjs},
         year = {2024a},
        month = dec,
       volume = {275},
       number = {2},
          eid = {36},
        pages = {36},
          doi = {10.3847/1538-4365/ad5ae2},
archivePrefix = {arXiv},
       eprint = {2303.01658},
 primaryClass = {astro-ph.GA},
       adsurl = {https://ui.adsabs.harvard.edu/abs/2024ApJS..275...36F}
}

@ARTICLE{2018ApJ...855....4K,
       author = {{Kawamata}, Ryota and {Ishigaki}, Masafumi and {Shimasaku}, Kazuhiro and {Oguri}, Masamune and {Ouchi}, Masami and {Tanigawa}, Shingo},
        title = "{Size-Luminosity Relations and UV Luminosity Functions at z = 6-9 Simultaneously Derived from the Complete Hubble Frontier Fields Data}",
      journal = {\apj},
         year = 2018,
        month = mar,
       volume = {855},
       number = {1},
          eid = {4},
        pages = {4},
          doi = {10.3847/1538-4357/aaa6cf},
archivePrefix = {arXiv},
       eprint = {1710.07301},
 primaryClass = {astro-ph.GA},
       adsurl = {https://ui.adsabs.harvard.edu/abs/2018ApJ...855....4K}
}

@ARTICLE{2021ApJ...908..192S,
       author = {{Sun}, Fengwu and {Egami}, Eiichi and {Rawle}, Timothy D. and {Walth}, Gregory L. and {Smail}, Ian and {Dessauges-Zavadsky}, Miroslava and {P{\'e}rez-Gonz{\'a}lez}, Pablo G. and {Richard}, Johan and {Combes}, Francoise and {Ebeling}, Harald and {Pell{\'o}}, Roser and {Van der Werf}, Paul and {Altieri}, Bruno and {Boone}, Fr{\'e}d{\'e}ric and {Cava}, Antonio and {Chapman}, Scott C. and {Cl{\'e}ment}, Benjamin and {Finoguenov}, Alexis and {Nakajima}, Kimihiko and {Rujopakarn}, Wiphu and {Schaerer}, Daniel and {Valtchanov}, Ivan},
        title = "{ALMA 1.3 mm Survey of Lensed Submillimeter Galaxies Selected by Herschel: Discovery of Spatially Extended SMGs and Implications}",
      journal = {\apj},
         year = 2021,
        month = feb,
       volume = {908},
       number = {2},
          eid = {192},
        pages = {192},
          doi = {10.3847/1538-4357/abd6e4},
archivePrefix = {arXiv},
       eprint = {2101.03677},
 primaryClass = {astro-ph.GA},
       adsurl = {https://ui.adsabs.harvard.edu/abs/2021ApJ...908..192S}
}

@INPROCEEDINGS{2023pcsf.conf...16K,
       author = {{Kohno}, K. and {Fujimoto}, S. and {Tsujita}, A. and {Kokorev}, V. and {Brammer}, G. and {Magdis}, G.~E. and {Valentino}, F. and {Laporte}, N. and {Sun}, Fengwu and {Egami}, E. and {Bauer}, F.~E. and {Guerrero}, A. and {Nagar}, N. and {Caputi}, K.~I. and {Caminha}, G.~B. and {Jolly}, J. -B. and {Knudsen}, K.~K. and {Uematsu}, R. and {Ueda}, Y. and {Oguri}, M. and {Zitrin}, A. and {Ouchi}, M. and {Ono}, Y. and {Gonz{\'a}lez-L{\'o}pez}, J. and {Richard}, J. and {Smail}, I. and {Coe}, D. and {Postman}, M. and {Bradley}, L. and {Koekemoer}, A.~M. and {Arancibia}, A.~M. Mu noz and {Dessauges-Zavadsky}, M. and {Espada}, D. and {Umehata}, H. and {Hatsukade}, B. and {Egusa}, F. and {Shimasaku}, K. and {Matsui-Morokuma}, K. and {Wang}, W. -H. and {Wang}, T. and {Ao}, Y. and {Baker}, A.~J. and {Lee}, Minju M. and {Lagos}, C. del P. and {Hughes}, D.~H. and {ALCS collaboration}},
        title = "{Unbiased surveys of dust-enshrouded galaxies using ALMA}",
    booktitle = {Physics and Chemistry of Star Formation: The Dynamical ISM Across Time and Spatial Scales},
         year = 2023,
       editor = {{Ossenkopf-Okada}, V. and {Schaaf}, R. and {Breloy}, I. and {Stutzki}, J.},
        month = feb,
        pages = {16},
          doi = {10.48550/arXiv.2305.15126},
archivePrefix = {arXiv},
       eprint = {2305.15126},
 primaryClass = {astro-ph.GA},
       adsurl = {https://ui.adsabs.harvard.edu/abs/2023pcsf.conf...16K}
}

@ARTICLE{2003PASP..115..763C,
       author = {{Chabrier}, Gilles},
        title = "{Galactic Stellar and Substellar Initial Mass Function}",
      journal = {\pasp},
         year = 2003,
        month = jul,
       volume = {115},
       number = {809},
        pages = {763-795},
          doi = {10.1086/376392},
archivePrefix = {arXiv},
       eprint = {astro-ph/0304382},
 primaryClass = {astro-ph},
       adsurl = {https://ui.adsabs.harvard.edu/abs/2003PASP..115..763C}
}

@ARTICLE{2023AA...680A..95Y,
       author = {{Yang}, Chentao and {Omont}, Alain and {Mart{\'\i}n}, Sergio and {Bisbas}, Thomas G. and {Cox}, Pierre and {Beelen}, Alexandre and {Gonz{\'a}lez-Alfonso}, Eduardo and {Gavazzi}, Rapha{\"e}l and {Aalto}, Susanne and {Andreani}, Paola and {Ceccarelli}, Cecilia and {Gao}, Yu and {Gorski}, Mark and {Gu{\'e}lin}, Michel and {Fu}, Hai and {Ivison}, R.~J. and {Knudsen}, Kirsten K. and {Lehnert}, Matthew and {Messias}, Hugo and {Muller}, Sebastien and {Neri}, Roberto and {Riechers}, Dominik and {van der Werf}, Paul and {Zhang}, Zhi-Yu},
        title = "{SUNRISE: The rich molecular inventory of high-redshift dusty galaxies revealed by broadband spectral line surveys}",
      journal = {\aap},
         year = 2023,
        month = dec,
       volume = {680},
          eid = {A95},
        pages = {A95},
          doi = {10.1051/0004-6361/202347610},
archivePrefix = {arXiv},
       eprint = {2308.07368},
 primaryClass = {astro-ph.GA},
       adsurl = {https://ui.adsabs.harvard.edu/abs/2023A&A...680A..95Y}
}

@ARTICLE{2018ApJ...869...27V,
       author = {{Valentino}, Francesco and {Magdis}, Georgios E. and {Daddi}, Emanuele and {Liu}, Daizhong and {Aravena}, Manuel and {Bournaud}, Fr{\'e}d{\'e}ric and {Cibinel}, Anna and {Cormier}, Diane and {Dickinson}, Mark E. and {Gao}, Yu and {Jin}, Shuowen and {Juneau}, St{\'e}phanie and {Kartaltepe}, Jeyhan and {Lee}, Min-Young and {Madden}, Suzanne C. and {Puglisi}, Annagrazia and {Sanders}, David and {Silverman}, John},
        title = "{A Survey of Atomic Carbon [C I] in High-redshift Main-sequence Galaxies}",
      journal = {\apj},
         year = 2018,
        month = dec,
       volume = {869},
       number = {1},
          eid = {27},
        pages = {27},
          doi = {10.3847/1538-4357/aaeb88},
archivePrefix = {arXiv},
       eprint = {1810.11029},
 primaryClass = {astro-ph.GA},
       adsurl = {https://ui.adsabs.harvard.edu/abs/2018ApJ...869...27V}
}

@ARTICLE{2013MNRAS.435.1493A,
       author = {{Alaghband-Zadeh}, S. and {Chapman}, S.~C. and {Swinbank}, A.~M. and {Smail}, Ian and {Danielson}, A.~L.~R. and {Decarli}, R. and {Ivison}, R.~J. and {Meijerink}, R. and {Weiss}, A. and {van der Werf}, Paul. P.},
        title = "{Using [C I] to probe the interstellar medium in z {\ensuremath{\sim}} 2.5 sub-millimeter galaxies}",
      journal = {\mnras},
         year = 2013,
        month = oct,
       volume = {435},
       number = {2},
        pages = {1493-1510},
          doi = {10.1093/mnras/stt1390},
archivePrefix = {arXiv},
       eprint = {1307.6593},
 primaryClass = {astro-ph.CO},
       adsurl = {https://ui.adsabs.harvard.edu/abs/2013MNRAS.435.1493A}
}

@ARTICLE{2020A&A...641A.155V,
       author = {{Valentino}, F. and {Daddi}, E. and {Puglisi}, A. and {Magdis}, G.~E. and {Liu}, D. and {Kokorev}, V. and {Cortzen}, I. and {Madden}, S. and {Aravena}, M. and {G{\'o}mez-Guijarro}, C. and {Lee}, M. -Y. and {Le Floc'h}, E. and {Gao}, Y. and {Gobat}, R. and {Bournaud}, F. and {Dannerbauer}, H. and {Jin}, S. and {Dickinson}, M.~E. and {Kartaltepe}, J. and {Sanders}, D.},
        title = "{CO emission in distant galaxies on and above the main sequence}",
      journal = {\aap},
         year = 2020,
        month = sep,
       volume = {641},
          eid = {A155},
        pages = {A155},
          doi = {10.1051/0004-6361/202038322},
archivePrefix = {arXiv},
       eprint = {2006.12521},
 primaryClass = {astro-ph.GA},
       adsurl = {https://ui.adsabs.harvard.edu/abs/2020A&A...641A.155V}
}

@ARTICLE{2015A&A...577A..50D,
       author = {{Dessauges-Zavadsky}, M. and {Zamojski}, M. and {Schaerer}, D. and {Combes}, F. and {Egami}, E. and {Swinbank}, A.~M. and {Richard}, J. and {Sklias}, P. and {Rawle}, T.~D. and {Rex}, M. and {Kneib}, J. -P. and {Boone}, F. and {Blain}, A.},
        title = "{Molecular gas content in strongly lensed z \raisebox{-0.5ex}\textasciitilde 1.5-3 star-forming galaxies with low infrared luminosities}",
      journal = {\aap},
         year = 2015,
        month = may,
       volume = {577},
          eid = {A50},
        pages = {A50},
          doi = {10.1051/0004-6361/201424661},
archivePrefix = {arXiv},
       eprint = {1408.0816},
 primaryClass = {astro-ph.GA},
       adsurl = {https://ui.adsabs.harvard.edu/abs/2015A&A...577A..50D}
}

@ARTICLE{2017A&A...605A..81D,
       author = {{Dessauges-Zavadsky}, M. and {Zamojski}, M. and {Rujopakarn}, W. and {Richard}, J. and {Sklias}, P. and {Schaerer}, D. and {Combes}, F. and {Ebeling}, H. and {Rawle}, T.~D. and {Egami}, E. and {Boone}, F. and {Cl{\'e}ment}, B. and {Kneib}, J. -P. and {Nyland}, K. and {Walth}, G.},
        title = "{Molecular gas properties of a lensed star-forming galaxy at z   3.6: a case study}",
      journal = {\aap},
         year = 2017,
        month = sep,
       volume = {605},
          eid = {A81},
        pages = {A81},
          doi = {10.1051/0004-6361/201628513},
archivePrefix = {arXiv},
       eprint = {1610.08065},
 primaryClass = {astro-ph.GA},
       adsurl = {https://ui.adsabs.harvard.edu/abs/2017A&A...605A..81D}
}

@ARTICLE{2016ApJ...833...67W,
       author = {{Walter}, Fabian and {Decarli}, Roberto and {Aravena}, Manuel and {Carilli}, Chris and {Bouwens}, Rychard and {da Cunha}, Elisabete and {Daddi}, Emanuele and {Ivison}, R.~J. and {Riechers}, Dominik and {Smail}, Ian and {Swinbank}, Mark and {Weiss}, Axel and {Anguita}, Timo and {Assef}, Roberto and {Bacon}, Roland and {Bauer}, Franz and {Bell}, Eric F. and {Bertoldi}, Frank and {Chapman}, Scott and {Colina}, Luis and {Cortes}, Paulo C. and {Cox}, Pierre and {Dickinson}, Mark and {Elbaz}, David and {G{\'o}nzalez-L{\'o}pez}, Jorge and {Ibar}, Edo and {Inami}, Hanae and {Infante}, Leopoldo and {Hodge}, Jacqueline and {Karim}, Alex and {Le Fevre}, Olivier and {Magnelli}, Benjamin and {Neri}, Roberto and {Oesch}, Pascal and {Ota}, Kazuaki and {Popping}, Gerg{\"o} and {Rix}, Hans-Walter and {Sargent}, Mark and {Sheth}, Kartik and {van der Wel}, Arjen and {van der Werf}, Paul and {Wagg}, Jeff},
        title = "{ALMA Spectroscopic Survey in the Hubble Ultra Deep Field: Survey Description}",
      journal = {\apj},
         year = 2016,
        month = dec,
       volume = {833},
       number = {1},
          eid = {67},
        pages = {67},
          doi = {10.3847/1538-4357/833/1/67},
archivePrefix = {arXiv},
       eprint = {1607.06768},
 primaryClass = {astro-ph.GA},
       adsurl = {https://ui.adsabs.harvard.edu/abs/2016ApJ...833...67W}
}

@ARTICLE{2021A&A...646A..76L,
       author = {{Loiacono}, Federica and {Decarli}, Roberto and {Gruppioni}, Carlotta and {Talia}, Margherita and {Cimatti}, Andrea and {Zamorani}, Gianni and {Pozzi}, Francesca and {Yan}, Lin and {Lemaux}, Brian C. and {Riechers}, Dominik A. and {Le F{\`e}vre}, Olivier and {B{\`e}thermin}, Matthieu and {Capak}, Peter and {Cassata}, Paolo and {Faisst}, Andreas and {Schaerer}, Daniel and {Silverman}, John D. and {Bardelli}, Sandro and {Boquien}, M{\'e}d{\'e}ric and {Burkutean}, Sandra and {Dessauges-Zavadsky}, Miroslava and {Fudamoto}, Yoshinobu and {Fujimoto}, Seiji and {Ginolfi}, Michele and {Hathi}, Nimish P. and {Jones}, Gareth C. and {Khusanova}, Yana and {Koekemoer}, Anton M. and {Lagache}, Guilaine and {Lubin}, Lori M. and {Massardi}, Marcella and {Oesch}, Pascal and {Romano}, Michael and {Vallini}, Livia and {Vergani}, Daniela and {Zucca}, Elena},
        title = "{The ALPINE-ALMA [C II] survey. Luminosity function of serendipitous [C II] line emitters at z {\ensuremath{\sim}} 5}",
      journal = {\aap},
         year = 2021,
        month = feb,
       volume = {646},
          eid = {A76},
        pages = {A76},
          doi = {10.1051/0004-6361/202038607},
archivePrefix = {arXiv},
       eprint = {2006.04837},
 primaryClass = {astro-ph.GA},
       adsurl = {https://ui.adsabs.harvard.edu/abs/2021A&A...646A..76L}
}
\bibliographystyle{aasjournal}



\end{document}